\documentclass[twocolumn]{aastex631}

\usepackage[T1]{fontenc}
\usepackage{ae,aecompl}

\usepackage{graphicx}	% Including figure files
\usepackage{amsmath}	% Advanced maths commands
\usepackage{amssymb}	% Extra maths symbols
\usepackage{booktabs,tabularx}
\usepackage{enumitem}
\usepackage{multirow}
\usepackage{hyperref}

\newcommand{\ee}[1]{\times 10^{#1} } 
\newcommand{\msun}{\text{M}_\odot}

\newcommand{\javelin}{\textsc{javelin}}
\begin{document}

\title{AGN STORM 2. VI. Mapping Temperature Fluctuations in the Accretion Disk of Mrk~817}
\shorttitle{AGN STORM 2: Mapping the Temperature Fluctuations of Mrk 817}
\shortauthors{Neustadt et al.}

%%%%%%%%%%%%%%%%%%%%%%%%%%%%%%%%%%%%%%%%%%%%%%%%%%%%%%%%%%%%%%%%%%%%%%%%%

\author[0000-0001-7351-2531]{Jack M. M. Neustadt}\thanks{Email: neustadt.7@osu.edu}
\affiliation{Department of Astronomy, The Ohio State University, 140 W. 18th Ave., Columbus, OH 43210, USA}

\author[0000-0001-6017-2961]{Christopher S. Kochanek}
\affiliation{Department of Astronomy, The Ohio State University, 140 W. 18th Ave., Columbus, OH 43210, USA}
\affiliation{Center for Cosmology and AstroParticle Physics, The Ohio State University, 191 West Woodruff Ave., Columbus, OH 43210, USA}

\author[0000-0001-5639-5484]{John Montano}
\affiliation{Department of Physics and Astronomy, 4129 Frederick Reines Hall, University of California, Irvine, CA, 92697-4575, USA}

\author[0000-0001-9092-8619]{Jonathan Gelbord}
\affiliation{Spectral Sciences Inc., 30 Fourth Ave, Suite 2, Burlington, MA 01803}

\author[0000-0002-3026-0562]{Aaron J. Barth}
\affiliation{Department of Physics and Astronomy, 4129 Frederick Reines Hall, University of California, Irvine, CA, 92697-4575, USA}

\author[0000-0003-3242-7052]{Gisella De~Rosa}
\affiliation{Space Telescope Science Institute, 3700 San Martin Drive, Baltimore, MD 21218, USA}

\author[0000-0002-2180-8266]{Gerard A. Kriss}
\affiliation{Space Telescope Science Institute, 3700 San Martin Drive, Baltimore, MD 21218, USA}

\author[0000-0002-8294-9281]{Edward M. Cackett}
\affiliation{Department of Physics and Astronomy, Wayne State University, 666 W. Hancock St, Detroit, MI, 48201, USA}

\author[0000-0003-1728-0304]{Keith Horne}
\affiliation{SUPA School of Physics and Astronomy, North Haugh, St.~Andrews, KY16~9SS, Scotland, UK}

\author[0000-0003-0172-0854]{Erin A. Kara}
\affiliation{MIT Kavli Institute for Astrophysics and Space Research, Massachusetts Institute of Technology, Cambridge, MA 02139, USA}

\author[0000-0001-8391-6900]{Hermine Landt}
\affiliation{Centre for Extragalactic Astronomy, Department of Physics, Durham University, South Road, Durham DH1 3LE, UK}

\author[0000-0002-6766-0260]{Hagai Netzer}
\affiliation{School of Physics and Astronomy and the Wise Observatory, Tel Aviv University, Tel Aviv 6997801, Israel}

%A
\author[0000-0003-2991-4618]{Nahum Arav}
\affiliation{Department of Physics, Virginia Tech, Blacksburg, VA 24061, USA}

%B
% \author[0000-0001-6301-570X]{\comment{Benjamin D. Boizelle}}
% \affiliation{Department of Physics and Astronomy, N284 ESC, Brigham Young University, Provo, UT, 84602, USA}

\author[0000-0002-2816-5398]{Misty C. Bentz}
\affiliation{Department of Physics and Astronomy, Georgia State University, 25 Park Place, Suite 605, Atlanta, GA 30303, USA}

% \author[0000-0001-5955-2502]{\comment{Thomas G. Brink}}
% \affiliation{Department of Astronomy, University of California, Berkeley, CA 94720-3411, USA}

% \author[0000-0002-1207-0909]{\comment{Michael S. Brotherton}}
% \affiliation{Department of Physics and Astronomy, University of Wyoming, Laramie, WY 82071, USA}

%C
% \author{\comment{Doron Chelouche}}
% \affiliation{Department of Physics, Faculty of Natural Sciences, University of Haifa, Haifa 3498838, Israel}
% \affiliation{Haifa Research Center for Theoretical Physics and Astrophysics, University of Haifa, Haifa 3498838, Israel}

%D
\author[0000-0001-9931-8681]{Elena Dalla Bont\`{a}}\thanks{Visiting Fellow at UCLan}
\affiliation{Dipartimento di Fisica e Astronomia ``G.  Galilei,'' Universit\`{a} di Padova, Vicolo dell'Osservatorio 3, I-35122 Padova, Italy}
\affiliation{INAF - Osservatorio Astronomico di Padova, Vicolo dell'Osservatorio 5 I-35122, Padova, Italy}
\affiliation{Jeremiah Horrocks Institute, University of Central Lancashire, Preston, PR1 2HE, UK}

\author[0000-0002-0964-7500]{Maryam Dehghanian}
\affiliation{Department of Physics, Virginia Tech, Blacksburg, VA 24061, USA}

\author[0000-0002-5830-3544]{Pu Du} 
\affiliation{Key Laboratory for Particle Astrophysics, Institute of High Energy Physics, Chinese Academy of Sciences, 19B Yuquan Road, Beijing 100049, People's Republic of China}

%E 
\author[0000-0001-8598-1482]{Rick Edelson} 
\affiliation{Eureka Scientific Inc., 2452 Delmer St. Suite 100, Oakland, CA 94602, USA}

%F
\author[0000-0003-4503-6333]{Gary J. Ferland}
\affiliation{Department of Physics and Astronomy, The University of Kentucky, Lexington, KY 40506, USA}

% \author[0000-0002-8224-1128]{\comment{Laura Ferrarese}}
% \affiliation{NRC Herzberg Astronomy and Astrophysics Research Centre, 5071 West Saanich Road, Victoria, BC, V9E 2E7, Canada}

\author[0000-0002-2306-9372]{Carina Fian}
\affiliation{Haifa Research Center for Theoretical Physics and Astrophysics, University of Haifa, Haifa 3498838, Israel}
\affiliation{School of Physics and Astronomy and the Wise Observatory, Tel Aviv University, Tel Aviv 6997801, Israel}

% \author[0000-0003-3460-0103]{\comment{Alexei V. Filippenko}}
% \affiliation{Department of Astronomy, University of California, Berkeley, CA 94720-3411, USA}
% \affiliation{Miller Institute for Basic Research in Science, University of California, Berkeley, CA 94720, USA}

\author[0000-0002-3365-8875]{Travis Fischer}
\affiliation{Space Telescope Science Institute, 3700 San Martin Drive, Baltimore, MD 21218, USA}

% \author[0000-0002-2445-5275]{\comment{Ryan J. Foley}}
% \affiliation{Department of Astronomy and Astrophysics, University of California, Santa Cruz, CA 92064, USA}

%G
\author[0000-0002-2908-7360]{Michael R. Goad}
\affiliation{School of Physics and Astronomy, University of Leicester, University Road, Leicester, LE1 7RH, UK}

\author[0000-0002-9280-1184]{Diego H. Gonz\'{a}lez Buitrago}
\affiliation{Instituto de Astronom\'{\i}a, Universidad Nacional Aut\'{o}noma de M\'{e}xico, Km 103 Carretera Tijuana-Ensenada, 22860 Ensenada B.C., M\'{e}xico}

\author[0000-0002-8990-2101]{Varoujan Gorjian}
\affiliation{Jet Propulsion Laboratory, M/S 169-327, 4800 Oak Grove Drive, Pasadena, CA 91109, USA}

\author[0000-0001-9920-6057]{Catherine J. Grier}
\affiliation{Department of Astronomy, University of Wisconsin-Madison, Madison, WI 53706, USA}

%H
\author[0000-0002-1763-5825]{Patrick B. Hall}
\affiliation{Department of Physics and Astronomy, York University, Toronto, ON M3J 1P3, Canada}

% \author[0000-0002-6733-5556]{\comment{Juan V. Hern\'{a}ndez Santisteban}}
% \affiliation{SUPA School of Physics and Astronomy, North Haugh, St.~Andrews, KY16~9SS, Scotland, UK}

\author[0000-0002-0957-7151]{Y. Homayouni}
\affiliation{Space Telescope Science Institute, 3700 San Martin Drive, Baltimore, MD 21218, USA}
\affiliation{Department of Astronomy and Astrophysics, The Pennsylvania State University, 525 Davey Laboratory, University Park, PA 16802}
\affiliation{Institute for Gravitation and the Cosmos, The Pennsylvania State University, University Park, PA 16802}

\author{Chen Hu}
\affiliation{Key Laboratory for Particle Astrophysics, Institute of High Energy Physics, Chinese Academy of Sciences, 19B Yuquan Road, Beijing 100049, People's Republic of China}

%I
\author[0000-0002-1134-4015]{Dragana Ili\'{c}}
\affiliation{Department of Astronomy, Faculty of Mathematics, University of Belgrade, Studentski trg 16,11000 Belgrade, Serbia}
\affiliation{Humboldt Research Fellow, Hamburger Sternwarte, Universit{\"a}t Hamburg, Gojenbergsweg 112, 21029 Hamburg, Germany}

%J

\author[0000-0003-0634-8449]{Michael D. Joner}
\affiliation{Department of Physics and Astronomy, N284 ESC, Brigham Young University, Provo, UT, 84602, USA}

%K
\author[0000-0001-5540-2822]{Jelle Kaastra}
\affiliation{SRON Netherlands Institute for Space Research, Niels Bohrweg 4, 2333 CA Leiden, The Netherlands}
\affiliation{Leiden Observatory, Leiden University, PO Box 9513, 2300 RA Leiden, The Netherlands}

\author[0000-0002-9925-534X]{Shai Kaspi}
\affiliation{School of Physics and Astronomy and the Wise Observatory, Tel Aviv University, Tel Aviv 6997801, Israel}

\author[0000-0003-0944-1008]{Kirk T. Korista}
\affiliation{Department of Physics, Western Michigan University, 1120 Everett Tower, Kalamazoo, MI 49008-5252, USA}

\author[0000-0001-5139-1978]{Andjelka B. Kova{\v c}evi{\'c}}
\affiliation{University of Belgrade-Faculty of Mathematics, Department of astronomy, Studentski trg 16 Belgrade, Serbia}

% \author[0000-0001-8638-3687]{\comment{Daniel Kynoch}}
% \affiliation{Astronomical Institute of the Czech Academy of Sciences, Boční II 1401/1a, CZ-14100 Prague, Czechia}

%L
\author[0000-0002-8671-1190]{Collin Lewin}
\affiliation{MIT Kavli Institute for Astrophysics and Space Research, Massachusetts Institute of Technology, Cambridge, MA 02139, USA}

\author[0000-0001-5841-9179]{Yan-Rong Li}
\affiliation{Key Laboratory for Particle Astrophysics, Institute of High Energy Physics, Chinese Academy of Sciences, 19B Yuquan Road, Beijing 100049, People's Republic of China}

%M
\author[0000-0002-0151-2732]{Ian M. McHardy}
\affiliation{School of Physics and Astronomy, University of Southampton, Highfield, Southampton SO17 1BJ, UK}

% \author[0000-0003-1081-2929]{\comment{Jacob N. McLane}}
% \affiliation{Department of Physics and Astronomy, University of Wyoming, Laramie, WY 82071, USA}

\author[0000-0002-4994-4664]{Missagh Mehdipour}
\affiliation{Space Telescope Science Institute, 3700 San Martin Drive, Baltimore, MD 21218, USA}

\author[0000-0001-8475-8027]{Jake A. Miller}
\affiliation{Department of Physics and Astronomy, Wayne State University, 666 W. Hancock St, Detroit, MI, 48201, USA}

% \author{\comment{Jake Mitchell}}
% \affiliation{Centre for Extragalactic Astronomy, Department of Physics, Durham University, South Road, Durham DH1 3LE, UK}

%N

%O

%P
\author{Christos Panagiotou}
\affiliation{MIT Kavli Institute for Astrophysics and Space Research, Massachusetts Institute of Technology, Cambridge, MA 02139, USA}

\author[0000-0003-1183-1574]{Ethan Partington}
\affiliation{Department of Physics and Astronomy, Wayne State University, 666 W. Hancock St, Detroit, MI, 48201, USA}

\author[0000-0002-2509-3878]{Rachel Plesha}
\affiliation{Space Telescope Science Institute, 3700 San Martin Drive, Baltimore, MD 21218, USA}

\author[0000-0003-1435-3053]{Richard W. Pogge}
\affiliation{Department of Astronomy, The Ohio State University, 140 W. 18th Ave., Columbus, OH 43210, USA}
\affiliation{Center for Cosmology and AstroParticle Physics, The Ohio State University, 191 West Woodruff Ave., Columbus, OH 43210, USA}

\author[0000-0003-2398-7664]{Luka \v{C}. Popovi\'{c}}
\affiliation{Astronomical Observatory, Volgina 7, 11060 Belgrade, Serbia}
\affiliation{Department of Astronomy, Faculty of Mathematics, University of Belgrade, Studentski trg 16,11000 Belgrade, Serbia}

\author[0000-0002-6336-5125]{Daniel Proga}
\affiliation{Department of Physics \& Astronomy, University of Nevada, Las Vegas, 4505 S. Maryland Pkwy, Las Vegas, NV, 89154-4002, USA}

%Q

%R
% \author[0000-0002-5359-9497]{\comment{Daniele Rogantini}}
% \affiliation{MIT Kavli Institute for Astrophysics and Space Research, Massachusetts Institute of Technology, Cambridge, MA 02139, USA}

%S
\author[0000-0003-1772-0023]{Thaisa Storchi-Bergmann}
\affiliation{Departamento de Astronomia - IF, Universidade Federal do Rio Grande do Sul, CP 150501, 91501-970 Porto Alegre, RS, Brazil}

\author[0000-0002-9238-9521]{David Sanmartim}
\affiliation{Rubin Observatory Project Office, 950 N. Cherry Ave., Tucson, AZ 85719, USA} 

\author[0000-0003-2445-3891]{Matthew R. Siebert}
\affiliation{Space Telescope Science Institute, 3700 San Martin Drive, Baltimore, MD 21218, USA}

\author[0000-0002-8177-6905]{Matilde Signorini}
\affiliation{Dipartimento di Fisica e Astronomia, Università di Firenze, via G. Sansone 1, 50019 Sesto Fiorentino, Firenze, Italy}
\affiliation{INAF - Osservatorio Astrofisico di Arcetri, Largo Enrico Fermi 5, I-50125 Firenze, Italy}
\affiliation{Department of Physics and Astronomy, University of California, Los Angeles, CA 90095, USA}

%T
% \author[0000-0002-8460-0390]{\comment{Tommaso Treu}}\thanks{Packard Fellow}
% \affiliation{Department of Physics and Astronomy, University of California, Los Angeles, CA 90095, USA}

%U

%V
\author[0000-0001-9191-9837]{Marianne Vestergaard}
\affiliation{Steward Observatory, University of Arizona, 933 North Cherry Avenue, Tucson, AZ 85721, USA}
\affiliation{DARK, The Niels Bohr Institute, University of Copenhagen, Jagtvej 155A, DK-2200 Copenhagen N, Denmark}

%W
% \author[0000-0001-9449-9268]{\comment{Jian-Min Wang}}
% \affiliation{Key Laboratory for Particle Astrophysics, Institute of High Energy Physics, Chinese Academy of Sciences, 19B Yuquan Road, Beijing 100049, People's Republic of China}
% \affiliation{School of Astronomy and Space Sciences, University of Chinese Academy of Sciences, 19A Yuquan Road, Beijing 100049, People's Republic of China}
% \affiliation{National Astronomical Observatories of China, 20A Datun Road, Beijing 100020, People's Republic of China}

% \author[0000-0003-1810-0889]{\comment{Martin J. Ward}}
% \affiliation{Centre for Extragalactic Astronomy, Department of Physics, Durham University, South Road, Durham DH1 3LE, UK}

% \author[0000-0002-5205-9472]{\comment{Tim Waters}}
% \affiliation{Department of Physics \& Astronomy, University of Nevada, Las Vegas, 4505 S. Maryland Pkwy, Las Vegas, NV, 89154-4002, USA}

%X

%Y

%Z
\author[0000-0003-0931-0868 ]{Fatima Zaidouni}
\affiliation{MIT Kavli Institute for Astrophysics and Space Research, Massachusetts Institute of Technology, Cambridge, MA 02139, USA}

\author[0000-0001-6966-6925]{Ying Zu}
\affiliation{Department of Astronomy, School of Physics and Astronomy, Shanghai Jiao Tong University, 800 Dongchuan Road, Shanghai, 200240, People's Republic of China}
\affiliation{Shanghai Key Laboratory for Particle Physics and Cosmology, Shanghai Jiao Tong University, Shanghai, 200240, People's Republic of China}

%%%%%%%%%%%%%%%%%%%%%%%%%%%%%%%%%%%%%%%%%%%%%%%%%%%%%%%%%%%%%%%%%%%%%%%%%

% Abstract of the paper
\begin{abstract}
We fit the UV/optical lightcurves of the Seyfert 1 galaxy Mrk~817 to produce maps of the accretion disk temperature fluctuations $\delta T$ resolved in time and radius. The $\delta T$ maps are dominated by coherent radial structures that move slowly ($v \ll c$) inwards and outwards, which conflicts with the idea that disk variability is driven only by reverberation.  Instead, these slow-moving temperature fluctuations are likely due to variability intrinsic to the disk.  We test how modifying the input lightcurves by smoothing and subtracting them changes the resulting $\delta T$ maps and find that most of the temperature fluctuations exist over relatively long timescales ($\sim$100s of days).  We show how detrending AGN lightcurves can be used to separate the flux variations driven by the slow-moving temperature fluctuations from those driven by reverberation.  We also simulate contamination of the continuum emission from the disk by continuum emission from the broad line region (BLR), which is expected to have spectral features localized in wavelength, such as the Balmer break contaminating the $U$ band.  We find that a disk with a smooth temperature profile cannot produce a signal localized in wavelength and that any BLR contamination should appear as residuals in our model lightcurves.  Given the observed residuals, we estimate that only $\sim$20\% of the variable flux in the $U$ and $u$ lightcurves can be due to BLR contamination.  Finally, we discus how these maps not only describe the data, but can make predictions about other aspects of AGN variability.  
\end{abstract}


%%%%%%%%%%%%%%%%%%%%%%%%%%%%%%%%%%%%%%%%%%%%%%%%%%%%%%%%%%%%%%%%%%%%%%%%%

%%%%%%%%%%%%%%%%%%%%%%%%%%%%%%%%%%%%%%%%%%%%%%%%%%%%%%%%%%%%%%%%%%%%%%%%%
\section{Introduction}\label{sec:intro}

Understanding the continuum variability of active galactic nuclei (AGNs) is fundamental to understanding the accretion process for supermassive black holes (SMBHs).  The stochastic nature of this variability has been studied for decades (e.g., \citealt{oknyanskij78,perola82,ulrich97,cristiani97,giveon99,geha03,kelly09,macleod10,kozlowski10,davis20,burke21}) and is thought to be caused by temperature fluctuations in the accretion disk surrounding the SMBH.  Because shorter wavelengths are generally observed to vary first with lags between wavelengths typical of the light travel time across a disk (e.g., \citealt{sergeev05,cackett07}), the variability is frequently described by a ``lamppost'' reverberation model \citep{krolik91}.  In this model, fluctuations in the luminosity of the central region illuminate the outer regions and drive temperature fluctuations in the disk which in turn drive the variability. 

This assumption is often used for disk reverberation mapping (disk RM) where the inter-band lags are used to constrain the temperature profile of the disk (e.g., \citealt{shappee14,fausnaugh16,edelson17,vincentelli21}).  This technique is similar to broad line reverberation mapping, which uses the continuum and emission line lightcurves to measure the light travel time between the accretion disk and the broad line region (BLR) \citep{blandford82,peterson93}.  Frequently, the variable central source in disk RM studies is the X-ray corona \citep{nayakshin00,frank02}, but, there are cases where the X-rays vary after the UV/optical or show uncorrelated structures that call this assumption into question (e.g., \citealt{berkley00,kazanas01,mchardy14,mchardy18,edelson19,dexter19,cackett20,hernandez20,kara23}).  In most studies of disk RM, the model is generally only invoked to measure the inter-band lags rather than to analytically relate the X-ray fluctuations and the UV/optical response, with some exceptions (e.g., \citealt{shappee14,kammoun21}).  

There are also multiple open questions in disk RM studies, one being possible ``contamination'' from the BLR, which is more physically extended than the disk and would thus have a longer lag signature.  The main evidence for this is a ``bump'' in the lag spectrum around the Balmer break (3645~\AA), with longer lags in bands that cover this wavelength, like Swift~$U$ and SDSS~$u$.  This bump has been observed for many AGNs \citep{edelson15,edelson17,edelson19,fausnaugh16,cackett18,cackett20,hernandez20} and successfully modeled using various BLR gas models \citep{korista01,korista19,lawther18,netzer20,netzer22}, but not every AGN has this bump \citep{mchardy23} -- including Mrk~817 \citep{kara21,cackett23}.

There have been studies that argue against the lamppost reverberation model.  For example, \citet{dexter11} argue that disk variability can be modeled with inhomogeneous and non-axisymmetric temperature fluctuations across the disk, although \citet{kokubo15} finds that this conflicts with the tight correlations between bands.  Others argue for thermal fluctuations in the disk driven by processes other than reverberation (e.g., \citealt{cai18,cai20,sun20a,sun20b,li21}). Statistical analyses of the variability have modeled the variability as a modest-amplitude damped random walk (DRW, \citealt{kelly09,kozlowski10,macleod10,macleod12,zu13}) and have found that the timescales of the DRW are typical of the thermal timescales at the disk radii producing the observed flux in a given band and that they correlate with the mass of the SMBH \citep{kelly09,macleod10,burke21}.  These long timescales are a significant problem for the reverberation model, as they are much longer than any characteristic timescale associated with the very inner regions of the disk.  

\citet{neustadt22} introduced a model of disk variability that tries to reconstruct the temperature fluctuations in time and radius by inverting the UV/optical lightcurves, to produce a map of the accretion disk.  The inversion makes several  assumptions, including that (a) the steady-state temperature profile of the disk is that of the standard \citet{shakura73} thin-disk model, (b) the temperature fluctuations are axisymmetric, and (c) the temperature fluctuations are small and relatively ``smooth.'' The authors applied the model to well-sampled, multi-band lightcurves of seven AGNs, including the AGN Space Telescope and Optical Reverberation Mapping (AGN STORM) data for NGC~5548 \citep{derosa15,edelson15,fausnaugh16,starkey17}.  They found that the majority of AGNs show strong evidence for coherent temperature fluctuations that move slowly ($v \ll c$) radially inwards and/or outwards in the disk.  This is in conflict with the idea that reverberation -- which produces fast ($v \sim c$) signals that only move radially outwards -- is the only driving mechanism of disk variability.

The slow-moving fluctuations do not dominate the lightcurves, even though they tend to have higher temperature amplitudes, because the width of the blackbody curve in wavelength space means that a broad range of radii contribute to any given band.  Fluctuations that move slowly through the disk, and thus perturb a narrow range of radii over a given timescale, are washed out in comparison to fast-moving fluctuations.  Because the reverberation signal moves at roughly the speed of light (e.g., \citealt{cackett21}), and thus perturbs a broad range of radii over a short timescale, it is always going to be the least suppressed and most prominent feature of the lightcurves, even if the scale of temperature fluctuations produced by the reverberation signal is smaller than the slower-moving fluctuations.  This does not depend on whether the signal is moving inwards or outwards -- it depends only on the speed of the signal (see Sec.~6 and Fig.~20 of NK22).  

\citet{stone23} used the NK22 model on a sample of Sloan Digital Sky Survey (SDSS, \citealt{ahumada20}) quasars that had been spectroscopically monitored for years as part of the SDSS Reverberation Mapping campaign (SDSS-RM, \citealt{shen15,shen19}).  Despite the dramatically different data set (\citealt{stone23} used time series of spectra while NK22 used multi-band lightcurves), different cadences, and different AGN properties (the SDSS-RM quasars are much more massive, more luminous, and higher redshift than the NK22 sample), \citet{stone23} found similar results -- the majority of the temperature maps are dominated by coherent, slow-moving, radial temperature fluctuations.  They also found little to no evidence for reverberation signals in their maps, although this is probably due to the slow observing cadence of the SDSS-RM data relative to the light travel time of the disk.  

NK22 pointed towards a possible physical mechanism that could drive these fluctuations in the form of inwardly-propagating viscosity fluctuations that in turn drive accretion fluctuations  (\citealt{lyubarskii97}, also \citealt{kotov01,arevalo06}). Indeed, these accretion fluctuations have been invoked in previous studies to explain the UV/optical and X-ray variability on timescales longer than reverberation (e.g., \citealt{arevalo08,arevalo09,breedt09}).  These viscosity-driven fluctuations are thought to move only inwards along the disk, whereas the maps from NK22 and \citet{stone23} show fluctuations moving both outwards and inwards.  Another explanation is that these fluctuations are driven by opacity conditions, particularly the iron opacity bump, that can produce strong variations in temperature and luminosity at the disk radii producing the observed flux and on timescales of $\sim$100s of days \citep{jiang19,jiang20}.

In this paper, we apply the NK22 approach to data from the AGN STORM~2 campaign targeting Mrk~817.  The AGN STORM~2 project is a large-scale spectroscopic and photometric reverberation mapping campaign using X-ray through near-infrared observations from space- and ground-based observatories.  Previous papers include an overview of the first 100~days of observations \citep{kara21}, an analysis of the UV spectra obtained with the Hubble Space Telescope (HST, \citealt{homayouni23a}), an analysis of the X-ray properties using the Neil Gehrels Swift Observatory and the Neutron Star Interior Composition ExploreR (Swift and NICER, \citealt{partington23}), an overview of the UV flux variability and disk reverberation signal using Swift \citep{cackett23}, an analysis of the anomalous behavior of the broad \ion{C}{4} emission line lightcurve \citep{homayouni23b}, and an overview of the ground-based optical observations (Montano et al. in prep.).  In Section~\ref{sec:methods}, we summarize our model and discuss the additional analyses we use for Mrk~817.  First, we smooth the lightcurves over various timescales to explore how the inferred temperature fluctuations change.  Second, we subtract these smoothed lightcurves from our original unsmoothed lightcurves to see how much signal is removed from the resulting temperature fluctuations.  Third, we insert signals mimicking those expected for BLR contamination of the continuum emission. In Section~\ref{sec:results}, we discuss the Mrk~817 temperature maps and the effects of our manipulation of the lightcurves on the features of the maps.  In Section~\ref{sec:discussion}, we review our results, explain how our analyses place limits on the contamination from BLR continuum emission, and discuss the potential predictive (rather than simply descriptive) powers of our model.
 
%%%%%%%%%%%%%%%%%%%%%%%%%%%%%%%%%%%%%%%%%%%%%%%%%%%%%%%%%%%%%%%%%%%%%%%%%

\begin{figure}
\includegraphics[width=\linewidth]{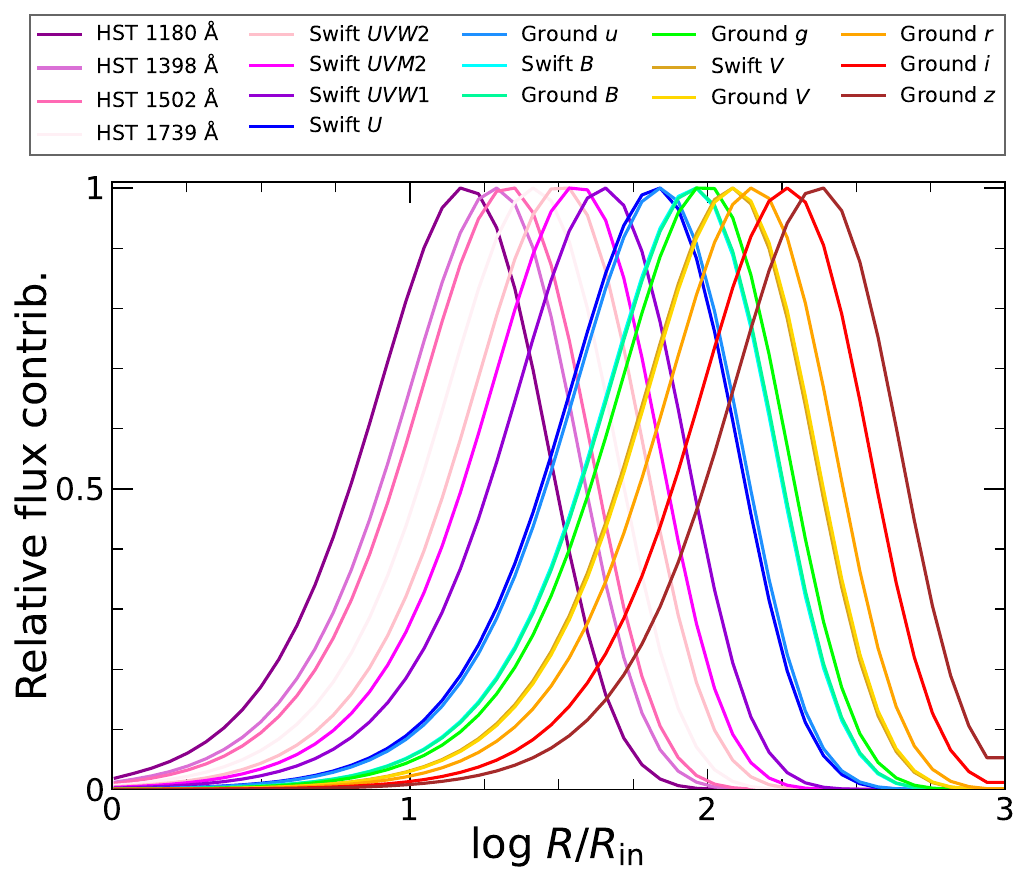}
\caption{Relative flux contribution from temperature fluctuations $\delta T$ to the AGN STORM~2 bands as a function for disk radius of Mrk~817.  The fluxes for each band are normalized to unity at peak.}
\label{fig:kernels}
\end{figure}

\section{Methods}\label{sec:methods}

\begin{figure*}
\centering
\includegraphics[width=0.99\linewidth]{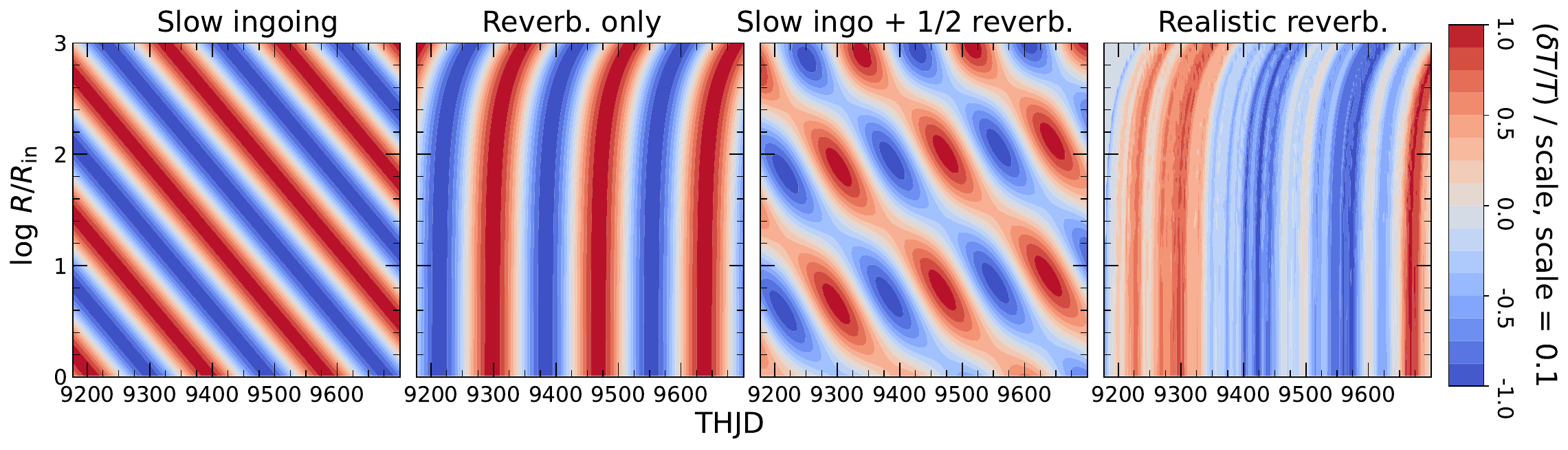}
\caption{Fractional $\delta T$ maps of different physical scenarios, from \textbf{left} to \textbf{right}: slow ingoing sinusoidal perturbations; an outgoing lamppost-like sinusoidal reverberation signal; a sum of the two where the ingoing signal is twice the strength of the outgoing reverberation signal; and an outgoing lamppost-like reverberation signal with a more ``realistic'' driving signal mimicking the Ground~$g$ lightcurve of Mrk~817.  The color scale spans the 99th percentile of the range of the $|\delta T/T|$ values, which here is fixed at 10\%.  A reverberation signal will look nearly vertical because a reverberation signal moves outward at the speed of light.  For Mrk~817, $\log R/R_{\rm in} = 2$ corresponds to 1.32 light-days ($3.41\ee{15}\rm~cm$).}
\label{fig:test_model}
\end{figure*}

The main equations that govern the model are detailed in NK22, but the most important parts of the model are explained and defined as follows. The model treats the lightcurve and the disk as two ``grids'' -- the lightcurves form a grid of fluxes in wavelength and time, and the disk is a grid of temperature fluctuations $\delta T$ in radius and time.  The model builds a system of linear equations that relates the grid elements of $\delta T$ to the corresponding fluxes, which are then inverted to fit the data and produce the grid or ``map'' of $\delta T$.  The steady-state temperature profile of the disk is assumed to be that of a \citet{shakura73} thin disk with $T(R)\propto R^{-3/4}$. The temperature fluctuations resolved in radius and time are $\delta T(R,t)$. There is no assumption about the physical mechanism driving $\delta T$.  The purpose of the model is to discern a physical mechanism based on the structures of the $\delta T$ maps, in particular how $\delta T$ propagates between radii and over time. A choice of a different $T(R)$ profile or emission profile, like one that accounts for optical depth effects (e.g., \citealt{pariev03}) does not change the qualitative structures of the $\delta T$ fluctuations so long as the radial temperature and emission profile is smooth. 

The disk is modelled using $N_R =50$ radial bins (this number is called $N_{u}$ in NK22) logarithmically spaced from $R/R_{\rm in} = 1$ to $R/R_{\rm in} = 1000$, where $R_{\rm in}$ is the inner radius of the disk.  We set $R_{\rm in} = 6 R_{g}$, where $R_g$ is the gravitational radius of the SMBH, and we assume the accretion efficiency $\eta = 0.1$, though this is large compared to the $\eta =0.056$ expected for a non-rotating BH \citep{laor89}.  Changing the value of $R_{\rm in}$ or $\eta$, like changing the mean temperature profile, also does not change the qualitative structures of the $\delta T$ fluctuations and only shifts the range of corresponding radii for each band.  The time dimension is divided into $N_t$ uniformly sampled intervals (this number is called $N_{t_p}$ in NK22).  In general, increasing $N_t$ leads to better fits with smaller values for the goodness of fit $\chi^2$, but little change in the qualitative structures of $\delta T$.  Larger $N_t$ also leads to larger computational costs, and so we chose $N_t = 250$ to match the approximate number of datapoints per band, though this does mean that the data sometimes have shorter cadences than the model.  In total, there are $N_d = 7380$ datapoints in the lightcurves compared to $N_R \times N_t = 12500$ points in the grid of the disk.

To create equations to transform flux variability into $\delta T$, the model assumes linear perturbations to the blackbody emission from the disk, thus requiring $\delta T/T$ to be small.  The effective contribution of $\delta T$ to the flux variability in each band as a function of radius is shown in Figure~\ref{fig:kernels}.  The inner radii contribute mostly to bluer bands, whereas the outer radii contribute mostly to redder bands, but there are also large radial overlaps between adjacent bands.  Note that the radial width of each band profile is completely dominated by the properties of the blackbody function -- the wavelength widths of the bands are not important.

In Figure~\ref{fig:test_model}, we show fractional $\delta T$ maps for several different physical scenarios: slow ingoing sinusoidal perturbations; an outgoing lamppost-like sinusoidal reverberation signal; a sum of the two where the ingoing signal is twice the strength of the outgoing reverberation signal; and an outgoing lamppost-like reverberation signal with a more ``realistic'' driving signal - rather than a sinusoid, the driving signal mimics the shape of the Ground~$g$ band lightcurve of Mrk~817.  A reverberation signal will look nearly vertical in these maps because a reverberation signal moves outward at the speed of light.  Note that these maps are not simulated or calculated using radiative transfer -- they are meant to reflect the general \textit{shapes} of the fluctuations in the maps that one might expect from the different physical scenarios. In NK22, we simulated observations based on these different scenarios and found that we are able to recover the input temperature fluctuations.

Due to the overlapping radial kernels, the finite temporal sampling, and temporal gaps in the data, the system of equations (see Eq. 10 in NK22) which must be inverted to construct $\delta T$ from the lightcurves is generically (nearly) degenerate.  The model of NK22 uses the technique of linear regularization, also called Tikhonov regularization, which adds in additional smoothing terms, to make the system of equations stably invertible.  Specifically, the model smooths over the overall scale of temperature fluctuations; the difference in $\delta T$ between adjacent radial bins, $\partial \delta T / \partial R$; and the difference in $\delta T$ between adjacent time bins, $\partial \delta T / \partial t$.  In linear regularization, the smoothing terms are weighted by a penalty factor $\xi$, also called the regularization parameter, and increasing $\xi$ has the effect of more heavily smoothing the resulting $\delta T$ map.  This also results in an array of temperature maps rather than a single map. There are multiple ways to choose an ``ideal'' $\xi$ value (see e.g, \citealt{press92,rezghi09,zhang10,edwards18,ivezic20}) which do not always agree, but we use these as references in evaluating an appropriate range of $\xi$ values.  We also discuss this briefly in Appendix~\ref{sec:append}.

To evaluate the results for different values of $\xi$, we consider the $\chi^2$ per datapoint $\chi^2/N_d$ and the scale of the $\delta T$ fluctuations. In the model, the $\chi^2$ is computed by inserting the output $\delta T$ fluctuations into the original system of equations transforming $\delta T$ to flux, producing a model lightcurve.  The $\chi^2$ is the difference between the real and model lightcurves weighted by the uncertainties.  The goodness of fit we use to evaluate our model is not the reduced $\chi^2_\nu$ (see App.~\ref{sec:append}).  In general, $\chi^2/N_d$ increases as $\xi$ increases -- the higher the smoothing, the worse the fits.  While this could lead one to pick the smallest possible $\xi$ to get $\chi^2/N_d \sim 1$, as is suggested in \citet{press92}, one needs to avoid overfitting.   Overfitting can be gauged by looking at the amplitude/scale of temperature fluctuations.  The scale of $\delta T$ increases with less smoothing (smaller $\xi$) and can easily reach $|\delta T/T| \sim 1$, producing nonphysical ``negative'' fluxes and violating our initial assumptions that the temperature fluctuations can be treated linearly.  For the simulated data models in NK22 (like those shown in Fig.~\ref{fig:test_model}) that mimicked the cadence, noise, and amplitudes of real observations (e.g., the AGN STORM~1 campaign), intermediate values of $\xi \sim 10$ to 100 reproduced the input fluctuations (this is shown in Sec.~3 of NK22).  For the rest of our analysis in this paper, we favor the solutions with $\xi=10$, but the qualitative structures in the $\delta T$ maps with $\xi = 1$ or $\xi=100$ are similar.

\subsection{The AGN STORM~2 Observations of Mrk~817}

We analyze the data for Mrk~817 from AGN STORM~2.  These data consist of a combination of photometry and spectroscopy using HST (DOI: 10.17909/n734-k698), Swift \citep{gehrels04,roming05}, and various ground-based observatories. The period of observations initially lasted from THJD~9175.7--9700.4 (2020 November 22 to 2022 April 30), where the Truncated HJD (THJD) is THJD = HJD -- 2450000.  Swift and the ground-based observatories observed nearly every day, while HST observed roughly every 2 days.  The observations and reductions are described in detail by \citet{homayouni23a} for HST, \citet{cackett23} for Swift, and \citet{kara21} and Montano et al. (in prep) for the ground-based data.  There are occasional gaps in the data across various bands due to spacecraft problems.  Ground-based imaging observations were obtained from several facilities: the Las Cumbres Observatory Global Telescope (LCOGT, \citealt{brown13}) located at McDonald Observatory in Texas, the Calar Alto Observatory in Spain, the Liverpool Telescope \citep{steele04} located on the island of La Palma in the Canary Islands, the Wise Observatory in Israel \citep{brosch08}, the Yunnan Observatory in China, and the Dan Zowada Memorial Observatory in New Mexico. The intercalibrations between the ground-based data from different observatories are detailed in Montano et al. (in prep).  The data span from 1180~\AA\ to 8897~\AA\ in 17 photometric bands\footnote{Throughout the paper, ``band'' is synonymous with ``filter'' in most cases, but because the HST data are not actually filter photometry but instead integrated fluxes from spectra, we use ``band.''} (see Fig.~\ref{fig:kernels}).  For the ground-based data, the various observatories use slightly different bands that are equivalent to, but not exactly, the typical Johnson-Cousins, Bessel, and SDSS bands, and so are referred to as ``Ground'' bands.  The physical parameters for Mrk~817, including the SMBH mass $M_{\rm BH}$, the luminosity distance $D_L$, redshift $z$, and the Eddington ratio $\lambda_{\rm Edd}$, are given in Table~\ref{tab:params}.  The inclination $i$ has little effect on our analysis and is fixed to 30$\degr$. 

\begin{table}
\centering
\caption{Physical parameters for Mrk~817 used our analysis}
\begin{tabular}{lcc} \toprule
Parameter & Unit & Value \\ \midrule \midrule
$M_{\rm BH}$ & $10^7~\msun$ & 3.85 \\ 
$D_L$ & Mpc & 136  \\ 
$z$ & -- & 0.031455  \\ 
$\lambda_{\rm Edd}$ & $L_{\rm bol}/L_{\rm Edd}$ & 0.2 \\ 
\bottomrule
\end{tabular}
\begin{flushleft}
\textit{Notes:} $M_{\rm BH}$, $\lambda_{\rm Edd}$, and $z$ are adopted from \citet{kara21}.  $D_L$ is calculated from redshift using \citet{wright06} assuming $\Lambda$CDM, $h_0 = 69.6$, $\Omega_m = 0.286$, and a flat Universe.
\end{flushleft}
\label{tab:params}
\end{table}

\begin{figure*}
\includegraphics[width=\linewidth]{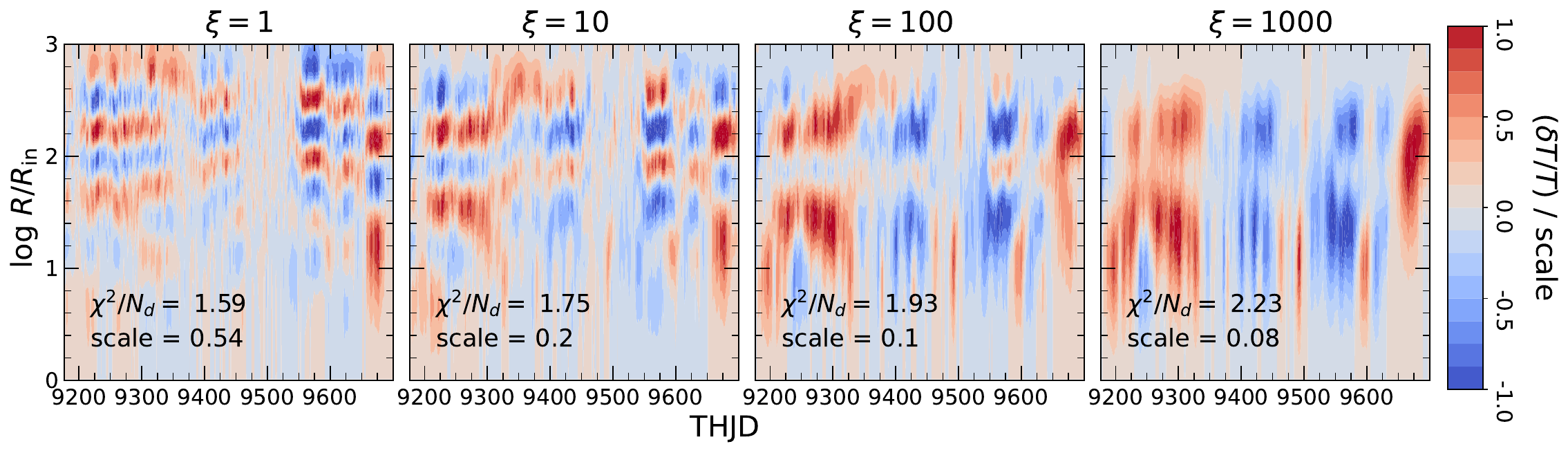}
\caption{Fractional $\delta T$ maps of Mrk~817 for smoothing parameters of $\xi = 1,10,100,1000$.  Each inversion gives the $\chi^2$ residuals per data point $\chi^2/N_d$ and the scale $|\delta T/T|$ for the colorbars.  The color scale spans the 99th percentile of the range of the $|\delta T/T|$ values.  Heavily smoothed maps (high $\xi$) look reverberation-like, whereas those with low or moderate smoothing (low or moderate $\xi$) show additional signals in the form of slow-moving ($v \ll c$) radial structures.  For the rest of our analysis, we use $\xi=10$ (see Sec.~\ref{sec:methods}).}
\label{fig:mrk817_model}
\end{figure*}

Following NK22, we check and correct for potential unaccounted systematic errors using a ``triplet test,'' where we fit each triplet of three adjacent lightcurve epochs with a line.  This assumes that the lightcurves are linear on timescales that are of the order of the cadence ($\sim$2~days for the HST data), which is reasonable given the timescales of AGN variability (10s of days, see, e.g., \citealt{burke21}). The $\chi_\nu^2$ for each set of three points should be 1 if the errors are correct, so we calculate the offset $\sigma$ that, when added to the reported errors in quadrature, makes $\chi_\nu^2=1$.  If this offset is negative, we use $\sigma = 0$, as we do not want to decrease the errors.  We then compute the median $\sigma$ per band and added this as a systematic increase to the errors in a given band.  Based on these tests, the errors for the Swift, ground-based, and HST~1709~\AA\ band lightcurves are left unchanged. The errors on the other, bluer HST lightcurves, which were calibrated by \citet{homayouni23a} to within 2\%, are inflated by a factor of 3.3--3.5.  While this seems large, the HST errors were inflated by similar values in \citet{cackett23} when they used \texttt{PyROA} \citep{donnan21} to measure the reverberation lags.  These rescalings numerically impact the resulting goodness-of-fit -- the $\chi^2/N_d$ changes to 1.75 from 2.27 for $\xi = 10$ -- but the error rescaling has little impact on the \textit{qualitative} structures in the resulting $\delta T$ maps. 

%%%%%%%%%%%%%%%%%%%%%%%%%%%%%%%%%%

\subsection{Manipulating lightcurves -- smoothing, subtracting, and inserting BLR signals}\label{sec:manipulate}

We examine how smoothing the lightcurves changes the resulting $\delta T$ maps.  This is different than the smoothing by the linear regularization parameter $\xi$.  Here, we are smoothing the lightcurves in time \textit{before} we perform the inversion with our model.  For the smoothing, we use Gaussians of various full-widths at half-maximums (FWHMs).  We do not change the errors of each datapoint.  The smoothing acts as a low-pass filter, only keeping the variability on timescales larger than the width.  If we subtract these smoothed lightcurves from the data, then we are left with the short timescale/high frequency variability which we can then invert with our model to see what these new maps imply for the temperature fluctuations on these short timescales.  This technique, also called ``detrending,'' is used in other reverberation studies of AGN lightcurves to separate reverberation signals on significantly different timescales (e.g., \citealt{welsh99,mchardy14,mchardy18,pahari20,lawther23}). 

While we do not expect contamination from the BLR emission lines to be important (see \citealt{zu11}), there are concerns about contamination from the BLR continuum emission.  To mimic BLR continuum contamination, we add an additional signal into the Swift~$U$ and Ground~$u$ (hereafter, $U$ and $u$) lightcurves, where the contribution from the BLR continuum is expected to be largest relative to that of the disk.  We take the raw lightcurve, and smooth it with a Gaussian with a FWHM of 10.5~days, which is half the approximate BLR lag of H$\beta$ for Mrk~817, measured to be $\sim$21~days by \citet{kara21}.  We then subtract the mean flux from this smoothed lightcurve so that we are only dealing with variable flux and not the steady-state flux. We use linear interpolation (\textsc{scipy.interpolate.interp1d}) to shift the smoothed, mean-subtracted lightcurve by a lag of 21~days,  multiply it by a scaling factor $f_{\rm BLR}$, and then add it into the original lightcurve.   The scaling factor is the fractional amplitude of the added signal compared to the variability in the original lightcurve.  This modeling is akin to treating the BLR as a uniform, face-on ring.  The values of $f_{\rm BLR}$ are chosen to reflect the possible fractional contamination by the BLR continuum, where $f_{\rm BLR} = 0.1$--0.5 corresponds to adding in an extra 10--50\% of variable flux into the $U$ and $u$ bands, the range predicted from BLR emission models \citep{korista01,korista19,netzer20,netzer22}.  The $U$ and $u$ lightcurves may already have contamination from the BLR continuum, but the purpose of this exercise is to see how the $\delta T$ maps change given a known level of BLR contamination and then use this to estimate the allowed level of contamination in the original lightcurves.

%%%%%%%%%%%%%%%%%%%%%%%%%%%%%%%%%%%%%%%%%%%%%%%%%%%%%%%%%%%%%%%%%%%%%%%%%

\section{Results}\label{sec:results}

\begin{figure*}
\includegraphics[width=\linewidth]{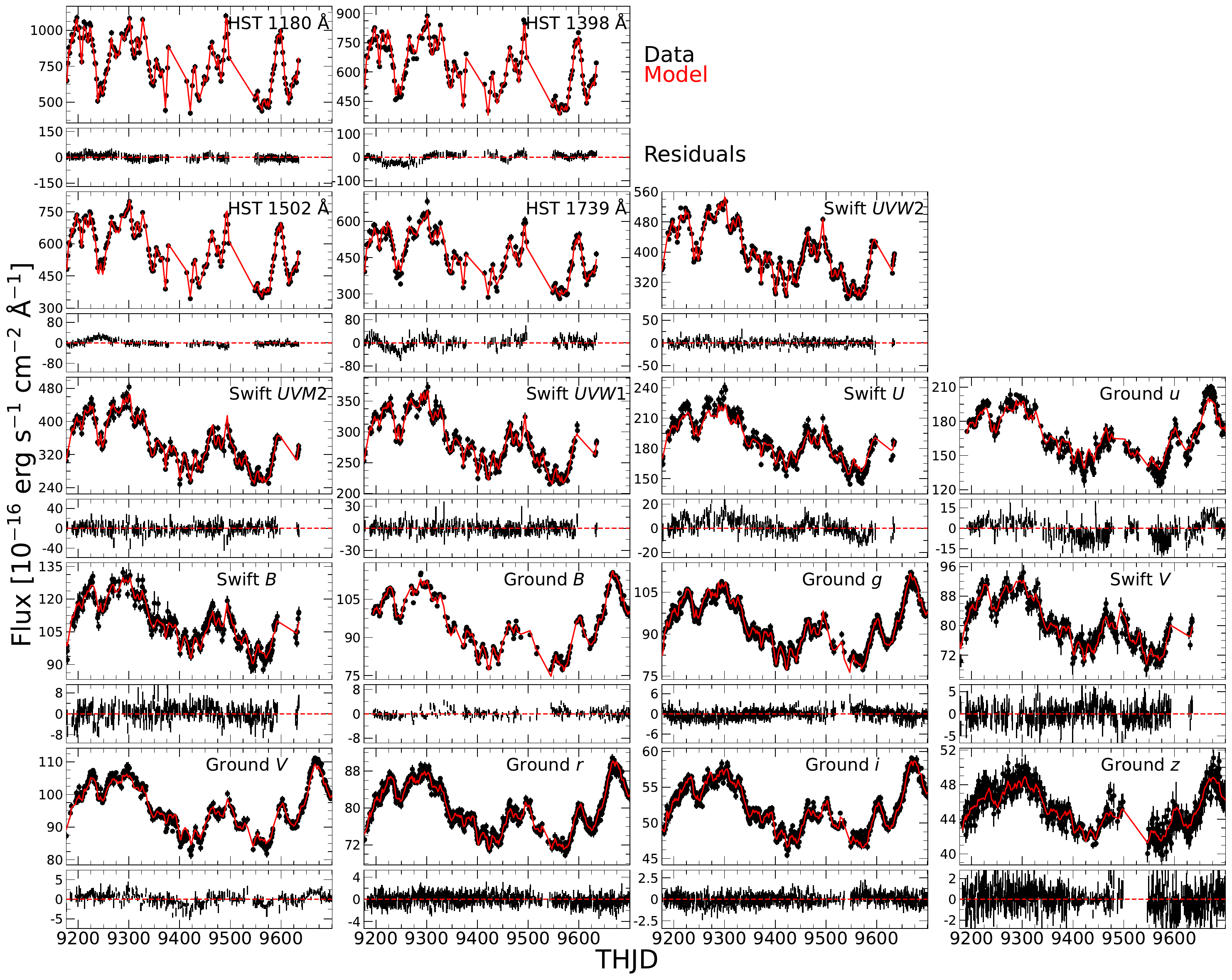}
\caption{Observed (black) and model (red) lightcurves and residuals of Mrk~817 with $\xi=10$  ($\chi^2/N_d = 1.75$, scale = 0.20).  There are some residual structures, most prominently in the $U$ lightcurve, which are analyzed in Sec.~\ref{sec:javelin}.}
\label{fig:mrk817_lc}
\end{figure*}

Figure~\ref{fig:mrk817_model} shows the fractional temperature fluctuation $\delta T$ maps for Mrk~817 using a range of $\xi$ values.  In Figure~\ref{fig:mrk817_lc}, we show the observed and model ($\xi=10$) lightcurves and the corresponding residuals. For the most part, the model fits the data quite well, though some bands (e.g., 1739~\AA, $U$, $u$) show clear structures in the residuals.  Whereas in NK22 we only used the residuals to compute the $\chi^2$, in this paper we will take a closer look at these residual structures (see Sec.~\ref{sec:javelin}).

\begin{figure}
\includegraphics[width=0.95\linewidth]{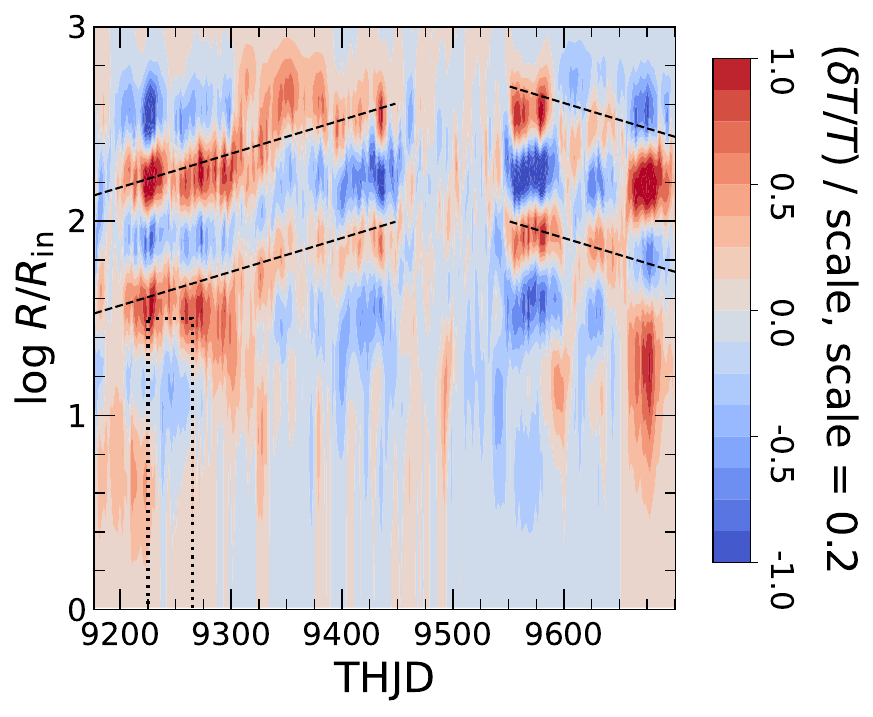}
\caption{Temperature maps of Mrk~817 with $\xi = 10$.  We highlight structures that strongly deviate from the reverberation model with dashed and dotted lines.  See Sec.~\ref{sec:results} for the discussion.}
\label{fig:mrk817_model_zoom}
\end{figure}

Examining Figure~\ref{fig:mrk817_model},  we see that there are prominent, coherent, radial fluctuations in the disk maps that appear to move slowly through the disks.  These fluctuations appear as alternating positive and negative radial structures that move together, though there are times when the structures disappear or become incoherent, like between THJD~9450 and 9550.  These features are similar to those observed for the AGNs modeled in NK22.  These structures are less prominent in the most highly smoothed maps ($\xi=1000$), but are arguably still present.  As we discuss in Section~\ref{sec:methods}, we disfavor the lowest and highest $\xi$ value solutions and favor the $\xi=10$ model as a good compromise between minimizing the $\chi^2/N_d$ and keeping the scale of $\delta T/T$ in the linear regime ($\lesssim 0.2$).  We isolate this model solution in Figure~\ref{fig:mrk817_model_zoom}.

In Figure~\ref{fig:mrk817_model_zoom}, we highlight the radial structures in the map that deviate strongly from the reverberation model, where the temperature fluctuations only move outwards at roughly the speed of light.  We mark out a particular feature in the dotted box - this is a negative temperature fluctuation that only exists inwards of $\log R/R_{\rm in} \sim 1.5$.  This appears to correspond to the ``dip'' in the lightcurves between THJD~9225 and 9275 that is more prominent in the bluer wavelengths (i.e., inner radii) than in the redder wavelengths.  As discussed in \citet{cackett23}, the first $\sim$150~days are not well modeled by a reverberation signal alone.  During this ``anomaly,'' the different bands are clearly not shifted and smoothed versions of a common signal.  A similar anomaly is detailed in \citet{homayouni23b}, where the broad UV emission line lightcurves are also not simply shifted and smoothed versions of the HST~1180~\AA\ band lightcurves.

We mark out the apparent motions of the main radial structures with dashed lines.  Before $\sim$THJD~9450, the apparent motion of the radial structures is outwards.  These radial structures seem to ``disappear'' or become totally incoherent after this date, though the structures reappear around $\sim$THJD~9550.  Interestingly, when we look at Figure~\ref{fig:mrk817_lc}, the period where the fluctuations are incoherent does not correspond to any obvious trends in the lightcurves, nor does it correspond to a lack of available data.  There is a gap in the HST~data between THJD~9500 and 9550, but this does not appear to strongly affect the maps.  Furthermore, this is 50~days after the fluctuations lose coherence around THJD~9450, implying that this is not driven by data availability or changes in data sampling. After $\sim$THJD~9550, the structures' apparent motions are more complicated, but overall appear to move inward.  

While the apparent motions are roughly linear on the maps, the radial scale is logarithmic, so the apparent velocity is increasing with radius, roughly as $v \propto R$, implying that the timescale associated with the fluctuations does not strongly depend on radius. This is also seen in the $\delta T$ maps in NK22 and \citet{stone23}.  Both of the dashed lines, inward and outward, correspond to physical velocities at $\log R/R_{\rm in}=2$ of $v \sim 1500 \rm ~km~s^{-1}$, which is roughly 13\% the orbital velocity at this radius and $0.005c$.  Similarly, the timescales of variations in $\delta T$ at a given radius do not significantly change with radius.  For example, the disk is not significantly more variable over time at $\log R/R_{\rm in}=1.5$ than at $\log R/R_{\rm in}=2.5$.  This also implies that the timescale associated with the temperature fluctuations does not strongly depend on radius.

This timescale is difficult to define, but we can make the following qualitative assessment of the map in Figure~\ref{fig:mrk817_model_zoom}.  If we look at the fluctuations between $1 \lesssim \log R/R_{\rm in} \leq 3$, the time taken to change from a negative to positive temperature fluctuation (or vice-versa) is $\sim$100~days.  This is purely determined by eye, focusing mostly on the fluctuations between THJD~9200 and 59450.  By comparison, the orbital timescale $t_{\rm orb}$ at $\log R/R_{\rm in}=2$ is $\sim$200~days.  We shall discuss the implications of this in Section~\ref{sec:discussion}.

Beyond $\sim$THJD~9650, there are no longer observations with HST or Swift from the initial campaign, and so the apparent radial structures beyond this date and below $\log R/R_{\rm in} \sim 1.5$ are not reliable.

%%%%%%%%%%%%%%%%%%%%%%%%%%%%%%%%%%

\subsection{Smoothed and subtracted lightcurves}\label{sec:smooth}

\begin{figure*}
\includegraphics[width=\linewidth]{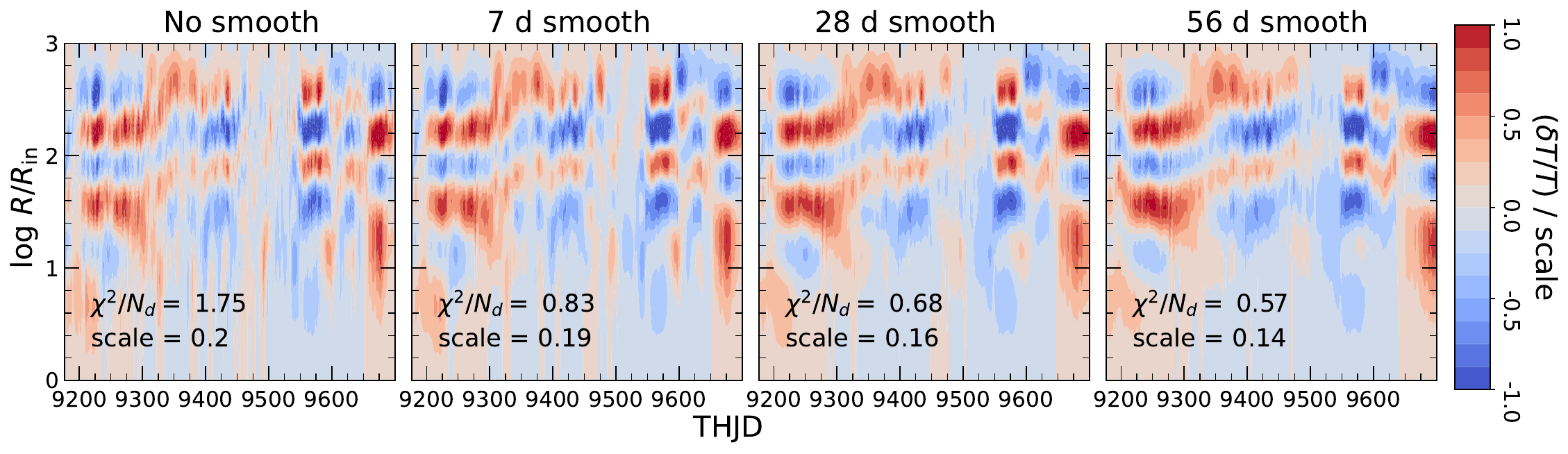}
\caption{Temperature maps ($\xi=10$) using the unsmoothed lightcurves (\textbf{left}) and using the lightcurves smoothed by Gaussians with FWHMs of 7, 28, and 56~days (\textbf{center left} to \textbf{right}).  The main differences between the original and smoothed models are that the small scale structures are suppressed (since they are effectively removed from the lightcurves) and that the $\chi^2/N_d$ significantly decreases.}
\label{fig:mrk817_smoothed}
\end{figure*}

\begin{figure*}
\includegraphics[width=\linewidth]{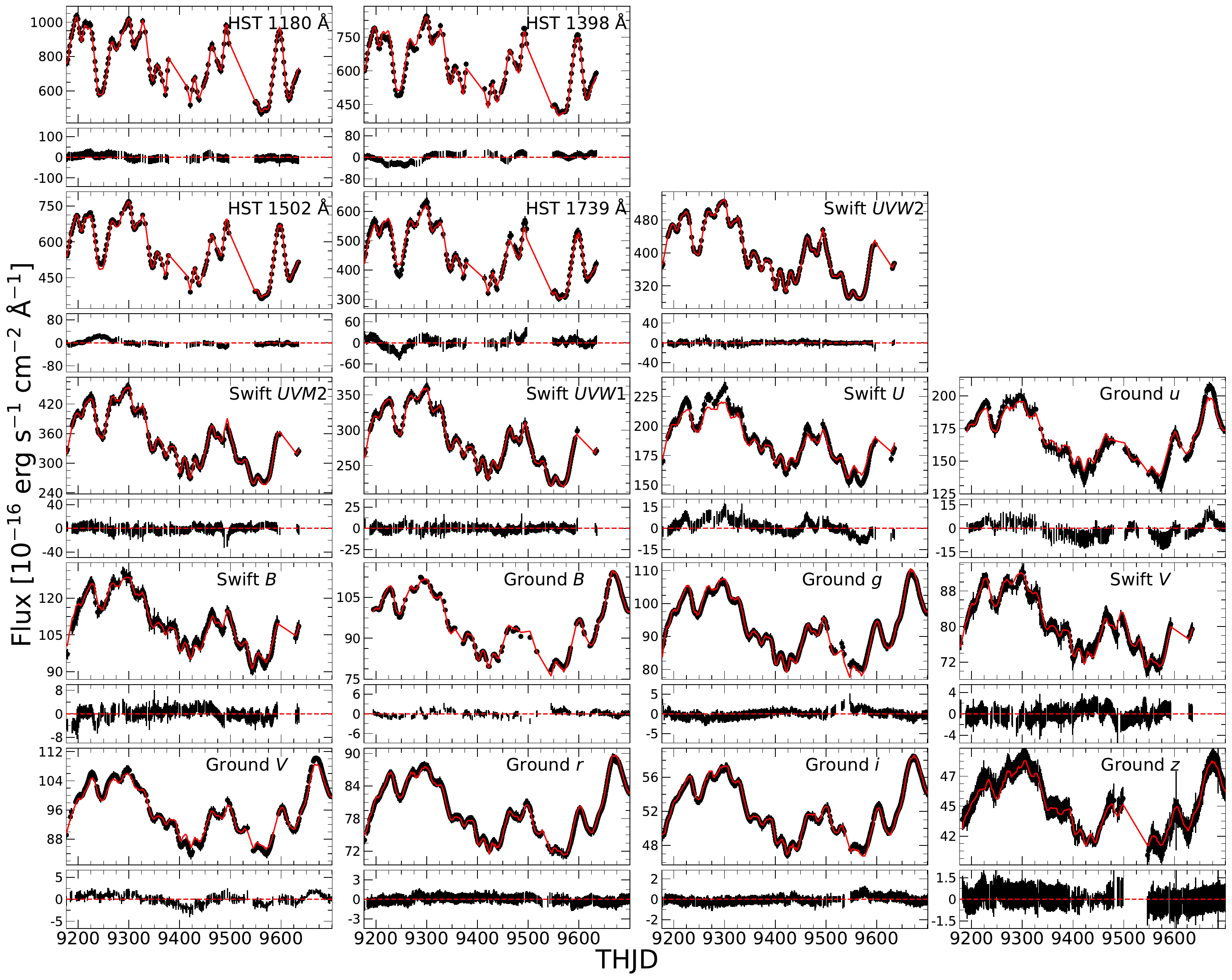}
\caption{Observed 7-day smoothed (black) and model (red, $\xi = 10$) lightcurves and residuals.}
\label{fig:mrk817_smoothed_lc}
\end{figure*}

In Figure~\ref{fig:mrk817_smoothed}, we show the temperature maps found by fitting the unsmoothed lightcurves and the lightcurves smoothed Gaussians with FWHMs of 7, 28, and 56~days.  In Figure~\ref{fig:mrk817_smoothed_lc}, we show the 7-day smoothed lightcurves, the model lightcurves, and the residuals.  As before, the model is constructed using $\xi=10$, and we do not change the errors of the data.  By comparing Figures~\ref{fig:mrk817_smoothed_lc} and \ref{fig:mrk817_lc}, we can see that short-term fluctuations in the lightcurves are effectively removed by the smoothing, as expected.  As we move from left to right (larger smoothing width) in Figure~\ref{fig:mrk817_smoothed}, the short-timescale structures steadily disappear.  The smoothing causes the $\chi^2$ to drop significantly, with the 7-day smoothed model having $\chi^2/N_d = 0.83$ instead of 1.75.  The $\chi^2$ continues to slightly decrease when increasing the smoothing to 28~days and then 56~days.  This exercise shows that all of the main radial structures we pointed out in our discussion of Figure~\ref{fig:mrk817_model_zoom} are independent of the lightcurve smoothing, and thus that the majority of the temperature fluctuations exist on relatively long timescales.

\begin{figure*}
\includegraphics[width=\linewidth]{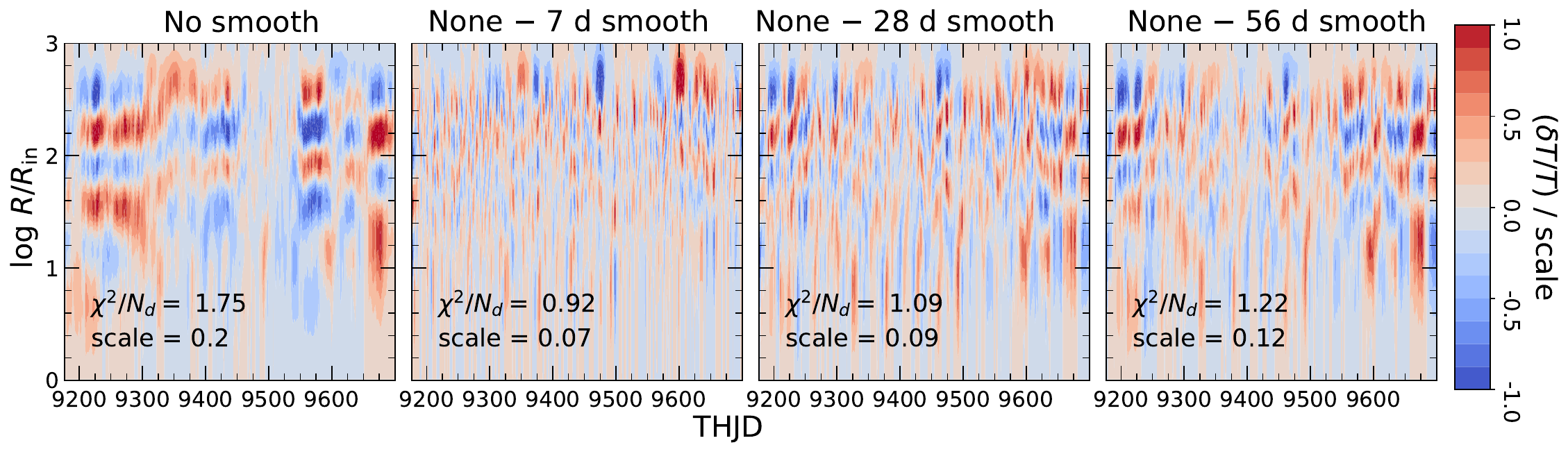}
\caption{Temperature maps ($\xi=10$) using the unsmoothed lightcurves (\textbf{left}) and using the lightcurves after subtracting the 7-, 28-, and 56-day smoothed lightcurves (\textbf{center left} to \textbf{right}).  There are no obvious large-timescale structures in the smoothed-and-subtracted maps, and instead these maps appear more dominated by short-timescale structures.}
\label{fig:mrk817_binsub}
\end{figure*}

\begin{figure*}
\includegraphics[width=\linewidth]{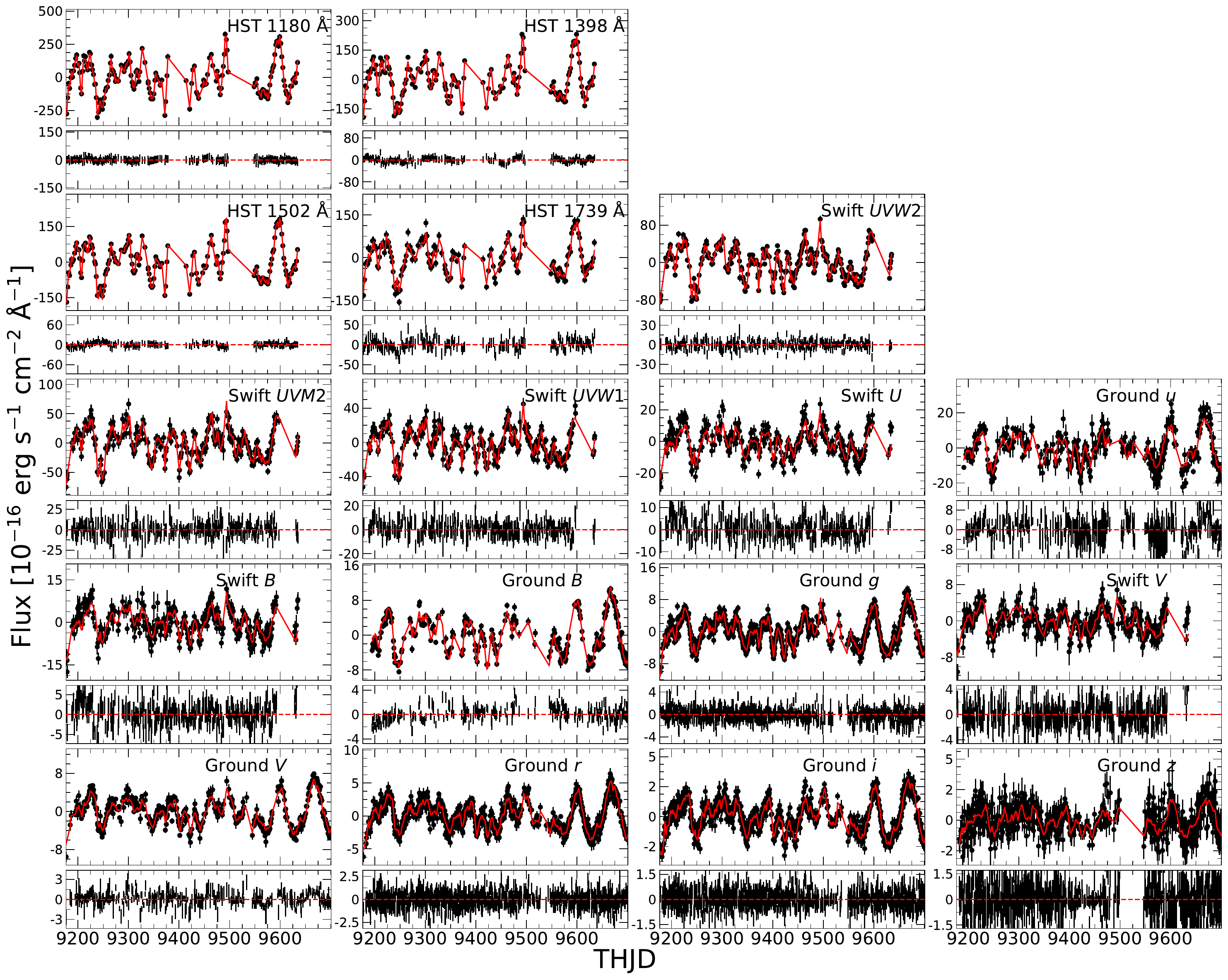}
\caption{Observed lightcurves subtracted by the 56-day smoothed lightcurves (black), the model lightcurves (red, $\xi=10$)  and the residuals.  Any structures in the residuals that are not also seen in the unsmoothed residuals (Fig.~\ref{fig:mrk817_lc}) are most likely due to edge effects and the large bin width. }
\label{fig:mrk817_binsub_lc}
\end{figure*}

%\textcolor{white}{Secret text to make sure that there isn't a stray line of text at the bottom of the page. Secret text to make sure that there isn't a stray line of text at the bottom of the page.}

We also model the lightcurves that have been subtracted by the smoothed lightcurves in order to examine the short-timescale variability. In Figure~\ref{fig:mrk817_binsub}, we show the temperature maps modeled using the unsmoothed/unsubtracted lightcurves and the lightcurves subtracted by the 7-, 28-, and 56-day smoothed lightcurves.  In Figure~\ref{fig:mrk817_binsub_lc}, we show the lightcurves with the 56-day smoothed lightcurve subtracted, the resulting model lightcurves, and the residuals.  As before, the model is constructed using $\xi=10$.  We see in Figure~\ref{fig:mrk817_binsub} that subtracting out the 7- and 28-day smoothed lightcurves from the original unsmoothed lightcurves removes almost all of the prominent large-timescale structures from the original temperature maps.  The remaining structures in the maps are more incoherent and only exist on short timescales ($\lesssim$ 20~days). Only for the 56-day smoothed-and-subtracted lightcurves is there some evidence for coherent structures, though they are not nearly as coherent as those seen in the original map.  We also cannot rule out that these structures are artifacts created by edge effects of the smoothing -- for example, the (arguably) most coherent fluctuations in the map are the inward-moving radial fluctuations after THJD~9500, but this is also a period with large gaps in the HST and Swift observations.  Overall, this experiment also shows that the majority of the temperature fluctuations exist over relatively long timescales. 

As discussed earlier in Section~\ref{sec:results} and in \citet{cackett23} and \cite{homayouni23b}, the first 150~days of HST and Swift data are ``anomalous'' in that the lightcurves do not look like smoothed and shifted versions of each other.  \citet{cackett23} addresses this by detrending the lightcurves with a $\sigma = 20$~days Gaussian (nearly equivalent to our FWHM = 56~days smoothing), where up to 50\% of the variability over long timescales is removed on the grounds that these long timescales are not the focus of the analysis.  Once detrended, the reverberation models are much better at matching the data, and the lag measurements more robustly converge.  In our analysis, the component that is excluded by the detrending corresponds to the slow-moving temperature fluctuations.  This is further evidenced by studies of the reverberation lags calculated from long-term ($\gtrsim$ 1~year) lightcurves, where detrending is sometimes \textit{required} to calculate the reverberation lag (see, e.g., \citealt{miller23}).  Similarly, in \citet{homayouni23b}, the mismatch between the UV emission lines and the 1180~\AA\ lightcurves is solved by adding a slowly-varying component to the emission line lightcurves, which implies a discrepancy between the 1180~\AA\ continuum and the extreme-UV (EUV) continuum that is driving the UV emission lines.  Using our framework, this discrepancy can be attributed to temperature fluctuations at the disk radii relevant for the 1180~\AA\ emission but not the EUV (or vice-versa).

\begin{figure*}
\includegraphics[width=\linewidth]{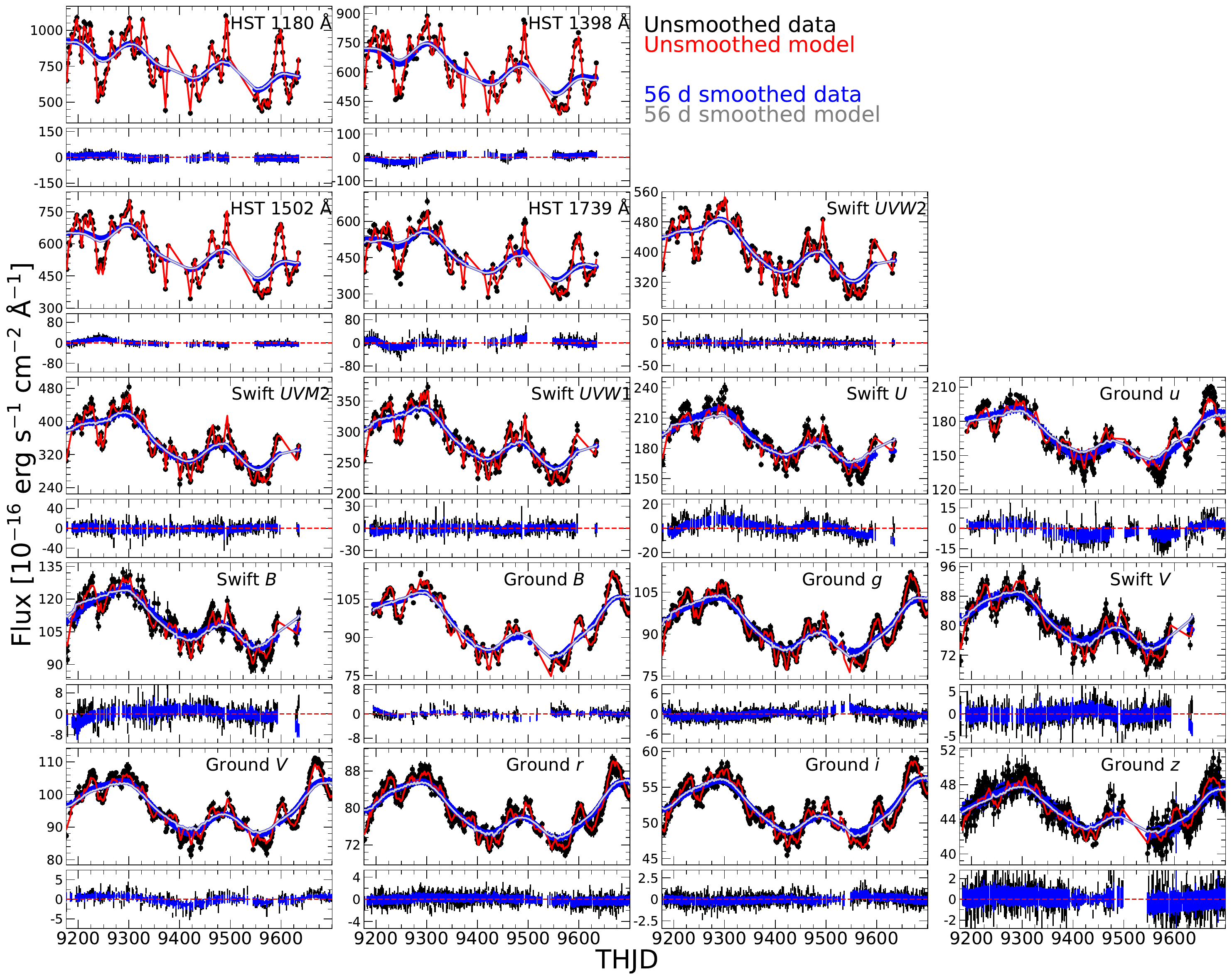}
\caption{Observed (black) and model (red) lightcurves and residuals of the unsmoothed lightcurves along with the observed (blue) and model (grey) lightcurves and residuals of the 56-day smoothed lightcurves. The lightcurves and residuals are normalized relative to the maxima/minima of the unsmoothed lightcurves.  One interpretation is that the smoothed lightcurves (blue and grey) track the slow-moving temperature fluctuations prominent in Figs.~\ref{fig:mrk817_model_zoom} and \ref{fig:mrk817_smoothed}, whereas unsmoothed lightcurves (black and red) show the additional effect of the reverberation signal as short-lived bumps and wiggles on top of the slow-moving perturbations.}
\label{fig:mrk817_detrend}
\end{figure*}

In Figure~\ref{fig:mrk817_detrend}, we show the unsmoothed and 56~d smoothed lightcurves and the model lightcurves from their corresponding temperature maps.  One can imagine this figure as separating the lightcurve into two components -- the long-timescale variability driven by the slow-moving temperature fluctuations in the disk (i.e., those highlighted in Fig.~\ref{fig:mrk817_model_zoom}), and the short-timescale variability driven by the reverberation.  This is not a perfect solution in that the remaining structures in the 56-day smoothed-and-subtracted model (rightmost panel in Fig.~\ref{fig:mrk817_binsub}) do not perfectly resemble lamppost-like fluctuations (see Fig.~\ref{fig:test_model}), but this still provides some clarity on the impact of the slow-moving temperature fluctuations on the lightcurves. 

%%%%%%%%%%%%%%%%%%%%%%%%%%%%%%%%%%

\subsection{Inserting and reproducing RM signals}\label{sec:rm}

\begin{figure*}
\includegraphics[width=\linewidth]{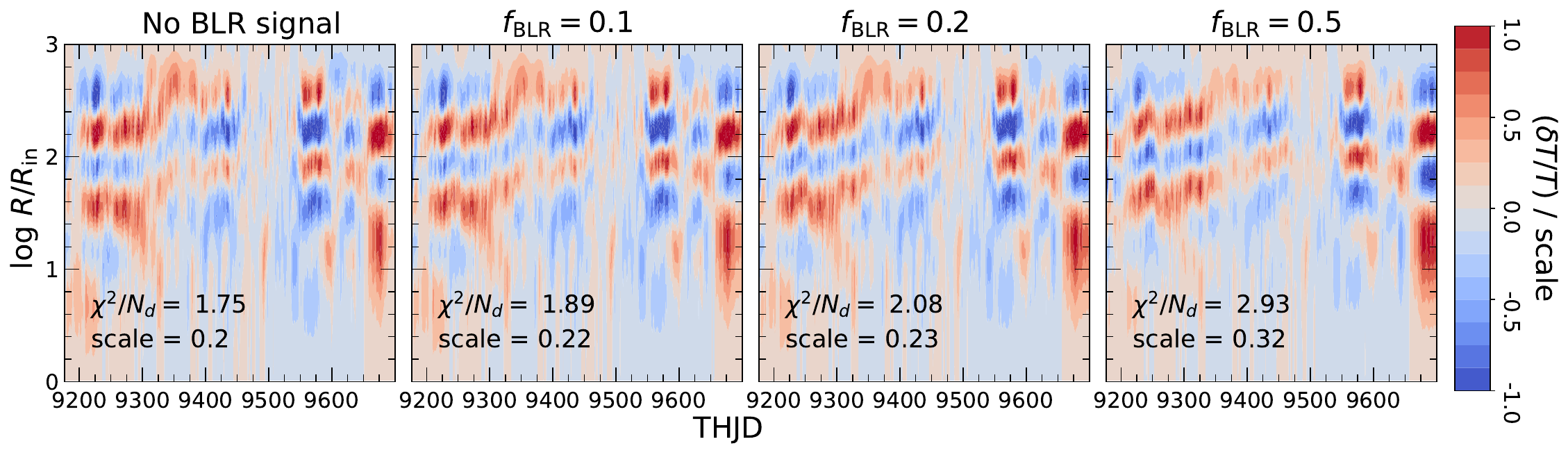}
\caption{Temperature maps ($\xi=10$) with no added BLR signal (\textbf{left}) and with the added BLR signal of various strengths $f_{\rm BLR} =$ 0.1, 0.2, and 0.5 (\textbf{center left} to \textbf{right}). The most prominent trends are that the maps do not change significantly, even with large $f_{\rm BLR}$, and that the $\chi^2$ and the scale of fluctuations both increase.}
\label{fig:mrk817_blr}
\end{figure*}

\begin{figure*}[htb!]
\includegraphics[width=\linewidth]{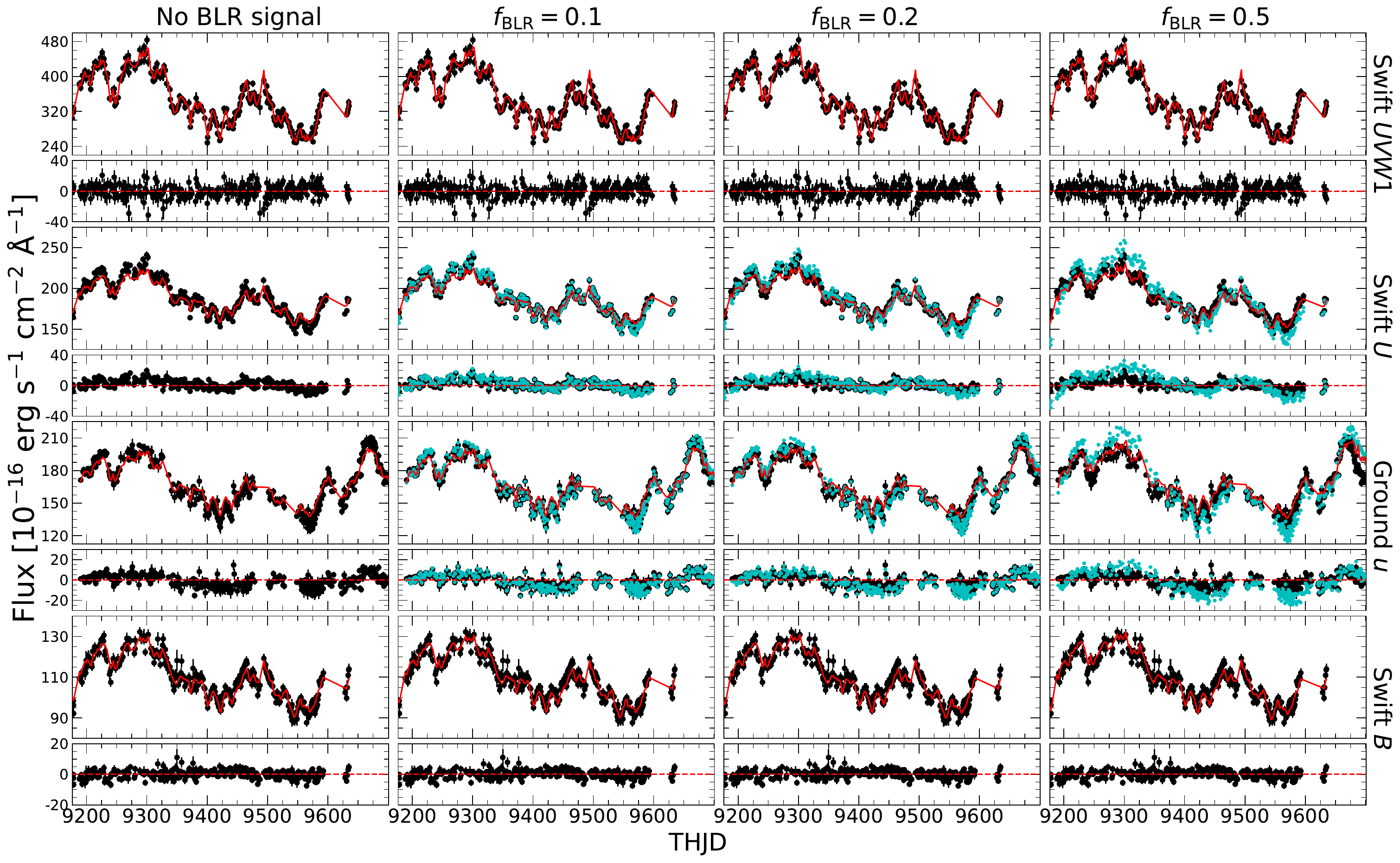}
\caption{Original (black), BLR-signal-added (turquoise), and model (red, $\xi=10$) lightcurves and residuals.  Because the BLR signal is only inserted into the $U$ and $u$ lightcurves (\textbf{middle} panels), we show only these and the closest bluer (Swift~\textit{UVW1}, \textbf{top} panel) and redder bands (Swift~$B$, \textbf{bottom} panel).  As we can see in the panels for the lightcurves with the inserted BLR signal -- the $U$ and $u$ -- the residuals do not change in structure with increasing $f_{\rm BLR}$, they only become more pronounced.  This is because our disk model cannot produce a signal localized in wavelength (see Sec.~\ref{sec:rm}).}
\label{fig:mrk817_blr_lc}
\end{figure*}

In Figure~\ref{fig:mrk817_blr}, we show the temperature maps constructed using the lightcurves with no added BLR signal and the lightcurves with a BLR signal added to the $U$ and $u$ lightcurves with fractional amplitudes of $f_{\rm BLR} = 0.1$, 0.2, and 0.5.  As described in Section~\ref{sec:manipulate}, $f_{\rm BLR} = 0.5$ means that the fluxes of the inserted $U$ and $u$ BLR signals are scaled to be 50\% of the variable flux of the respective lightcurves. In Figure~\ref{fig:mrk817_blr_lc}, we show the original lightcurves, the new lightcurves with the BLR signal inserted, the resulting model lightcurve, and the residuals.  Because the BLR signal is only inserted into the $U$ and $u$  lightcurves, we show only these and the closest bluer (Swift~\textit{UVW1}) and redder (Swift~$B$) band lightcurves.  As before, the model is constructed using $\xi=10$.

\begin{figure*}[htb!]
\includegraphics[width=1.00\linewidth]{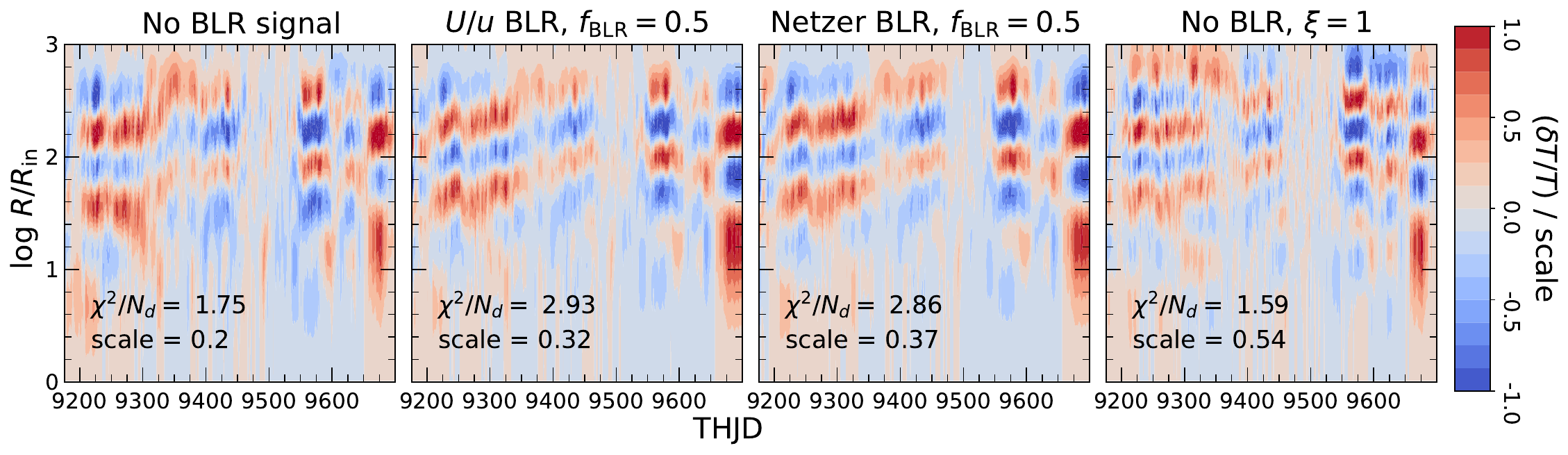}
\caption{Temperature maps ($\xi=10$) with no added BLR signal (\textbf{left}), with the BLR signal ($f_{\rm BLR} =0.5$) added only into the $U$ and $u$ lightcurves (\textbf{center left}), with the BLR signal added into every band according to the model based on \citet{netzer22} (\textbf{center right}), and no added BLR signal with a different smoothing parameter ($\xi=1$, \textbf{right}).  The $f_{\rm BLR} =0.5$ for the \citet{netzer22} model corresponds to the scale of fluctuations in $U$, where fluctuations in other bands are smaller as determined by the model.  The temperature maps of the $U$ and $u$ and the \citet{netzer22} scenarios are virtually identical - this is because there is still a significant difference in the scale of fluctuations between the $U$ and $u$ bands and the adjacent bands, which is not reproducible with the smooth temperature profile of the disk.  We also show how the maps with the inserted BLR signals mimic the maps without the signal with a smaller regularization parameter $\xi$ due to the increased $\chi^2$.}
\label{fig:mrk817_netzer}
\end{figure*}

Figure~\ref{fig:mrk817_blr} shows that the maps do not change significantly even after inserting a signal with $f_{\rm BLR} = 0.5$, although the $\chi^2$ and the scale of temperature fluctuations both increase. Interestingly, the model with $f_{\rm BLR} = 0.5$ somewhat resembles the $\xi=1$ model from Figure~\ref{fig:mrk817_model}, with less temporal and radial smoothing and with a larger scale of fluctuations (this is also shown later in Fig.~\ref{fig:mrk817_netzer}).  This is likely not a coincidence, since our model balances the $\chi^2$ and the smoothing in constructing the temperature map.  Having a larger overall $\chi^2$ causes the smoothing terms to be less important in the reconstruction, and thus the model is ``less smoothed'' for fixed $\xi$.   

In Figure~\ref{fig:mrk817_blr_lc}, we can see why the maps did not significantly change in Figure~\ref{fig:mrk817_blr} as a larger BLR signal was added and why the $\chi^2$ increased: the model lightcurves do not model the inserted flux.  As we can see in the panels for the bands with the inserted BLR signal -- the $U$ and $u$ bands -- the residuals do not change in structure with increasing $f_{\rm BLR}$, they only become more pronounced.  The reason for this can be seen in Figure~\ref{fig:kernels}.  Because each band receives flux contributions from a large range of radii, the temperature fluctuations needed to model a signal added to one band affects many other bands.  In other words, because of the large radial overlaps of the bands, temperature fluctuations at a given radius contributes to the flux in a wide range of wavelengths and thus in many bands. Our simulated BLR signal is wavelength-localized and only inserted into the $U$ and $u$ lightcurves, and as a result, the model cannot create temperature fluctuations at some radii without also producing flux in nearby bands. Thus, this inserted signal is ``ignored'' by the model.  To be clear, this is not a modelling problem -- a disk with a smooth, blackbody-like emission profile cannot produce a signal narrowly concentrated in wavelength.

We also model a more ``realistic'' BLR signal by adding the signal into each band lightcurve, not just the $U$ and $u$ bands, where the relative contribution of the signal to each band is set by a generic disk-BLR model based on \citet{netzer22}.  This model still has the BLR contribution peaking near the Balmer jump, but it also produce small amounts of flux at all wavelengths and thus all bands.  This model, along with the ``crude'' model that only includes the signal in the $U$ and $u$ bands, is shown in Figure~\ref{fig:mrk817_netzer}.  Here, $f_{\rm BLR} =0.5$ means that the scale of the $U$ band signal is 0.5, and the scale in other bands is scaled relative to this according to the model based on \citet{netzer22}.  Despite the BLR signal no longer being localized in wavelength, we find no noticeable differences in our reconstructed temperature maps between the more realistic model and our $U$- and $u$-only model. Even though some signal is now present in the other bands, the signal is still significantly stronger in the $U$ and $u$ bands than in adjacent bands and thus not reproducible with a disk.  Adding a wavelength dependent lag to the BLR contamination, as predicted by BLR models, would likely exacerbate these problems.

%%%%%%%%%%%%%%%%%%%%%%%%%%%%%%%%%%

\subsection{Measuring RM lags in the residuals}\label{sec:javelin}

\begin{figure*}
\includegraphics[width=\linewidth]{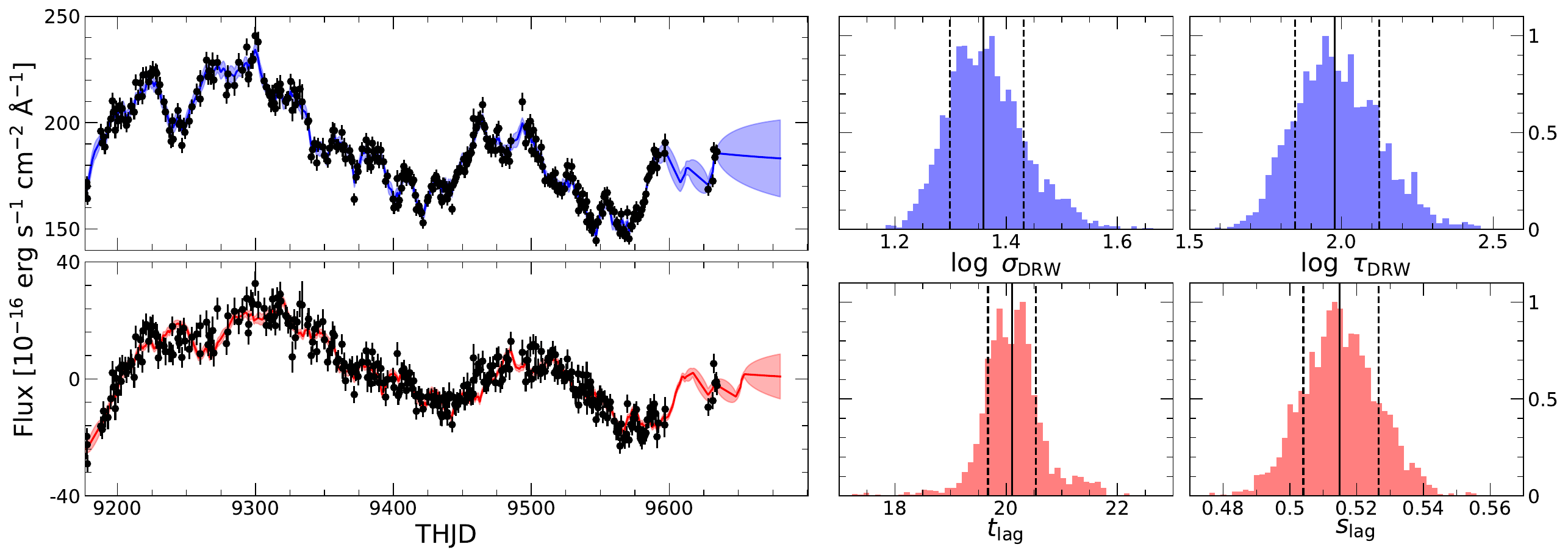}
\caption{\textbf{Left:} \javelin\ model fits to the unmodified $U$ lightcurve without an inserted BLR signal (\textbf{top}) and the residuals of a modeled lightcurve with an inserted signal of $f_{\rm BLR} = 0.5$ -- i.e., the residuals seen in Fig.~\ref{fig:mrk817_blr_lc} (\textbf{bottom}).  \textbf{Right:} Posterior distribution for the \javelin\ parameter fits, along with the median (solid line) and 1$\sigma$ confidence intervals (dashed line).  The recovered lag $t_{\rm lag}$ and scale $s_{\rm lag}$ parameters are almost identical to those of the inserted signal.}
\label{fig:mrk817_blr_javelin}
\end{figure*}

Even without inserting a BLR signal, there are clear structures in the $U$ and $u$ residuals seen in Figure~\ref{fig:mrk817_blr_lc} (and also Fig.~\ref{fig:mrk817_lc}).  These residual structures are especially coherent in the $U$ band.  We examine these residuals further and see if these are signals of BLR contamination by treating them as a RM problem and modeling them with the Python code \javelin\ \citep{zu11}.  As a consistency check, we run \javelin\ on our $U$ band residual lightcurve with an added $f_{\rm BLR} = 0.5$ signal, using the original $U$ lightcurve without the BLR signal as the ``driving continuum'' lightcurve.  \javelin\ models the continuum as a DRW and scales, shifts, and smooths the DRW to fit the lag signal. We show the resulting model lightcurves, including the DRW parameters, the damping timescale $\tau_{\rm DRW}$ and flux variability $\sigma_{\rm DRW}$ and the inferred lag parameters, lag time $t_{\rm lag}$ and lag scale $s_{\rm lag}$, in Figure~\ref{fig:mrk817_blr_javelin}.  We recover median values $t_{\rm lag} = 20.1^{+0.4}_{-0.4}$~days and $s_{\rm lag} = 0.52^{+0.01}_{-0.02}$, which are almost exactly the inserted lag of 21~days and BLR scaling $f_{\rm BLR} = 0.5$ within measurement uncertainties.  The lag smoothing ``width'' $w_{\rm lag}$ is relatively unimportant for our analysis, but hovers between 0 (no smoothing) and 1 (smoothed by 1~day) in all of the fits. 

\begin{figure*}
\includegraphics[width=\linewidth]{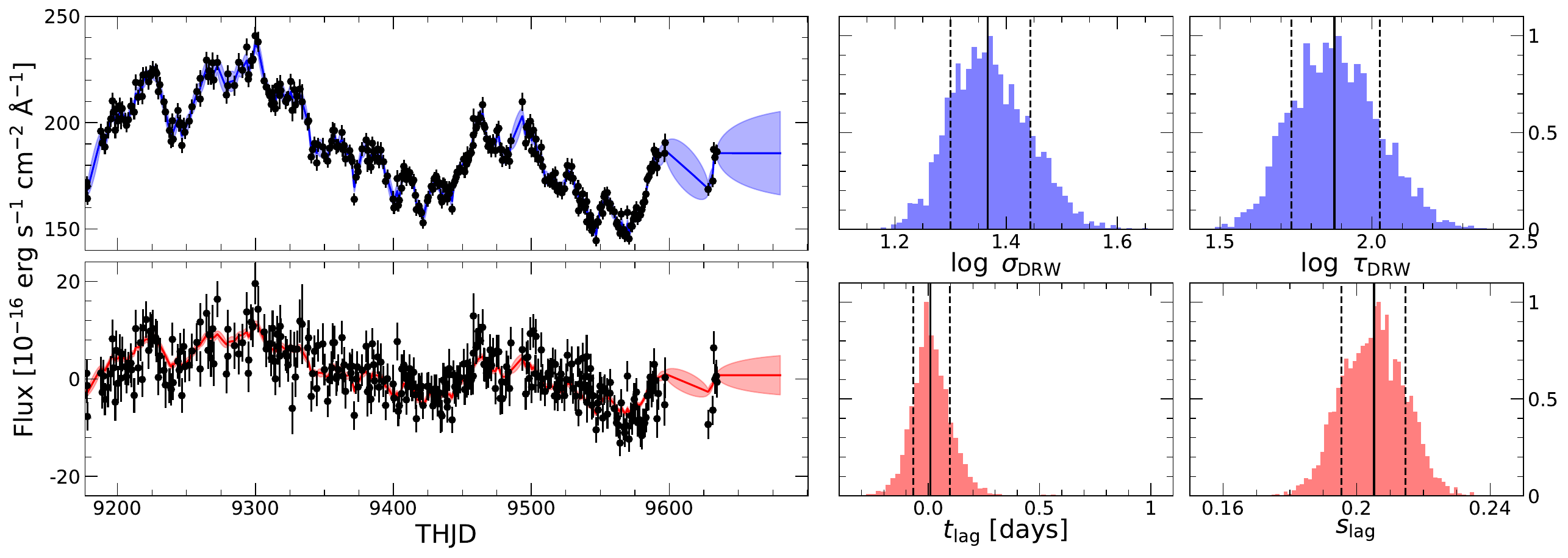}
\caption{\textbf{Left:} \javelin\ model fits to the unmodified $U$ lightcurve (\textbf{top}) and its corresponding residuals (\textbf{bottom}). \textbf{Right:} Posterior distribution for the \javelin\ parameter fits.  The recovered lag is consistent with 0~d, which is peculiar if this is extra flux due to BLR contamination.}
\label{fig:mrk817_resid_javelin}
\end{figure*}

Next, we repeat our analysis using the residual lightcurve with no inserted BLR signal.  We show the resulting fits in Figure~\ref{fig:mrk817_resid_javelin}.  Interestingly, this yields a lag of $t_{\rm lag} = 0.05^{+0.15}_{-0.10}$~days and a scale $s_{\rm lag} = 0.20^{+0.01}_{-0.01}$.  A lag consistent with 0~days is peculiar if the residuals are contamination from the BLR, but it is possible that this contamination originates from another source closer to the disk than the BLR that we will discuss later in the text. A scale of $\sim$0.2 implies that this extra flux accounts for $\sim$20\% of the flux variability in the $U$ band.  

\begin{figure*}
\includegraphics[width=\linewidth]{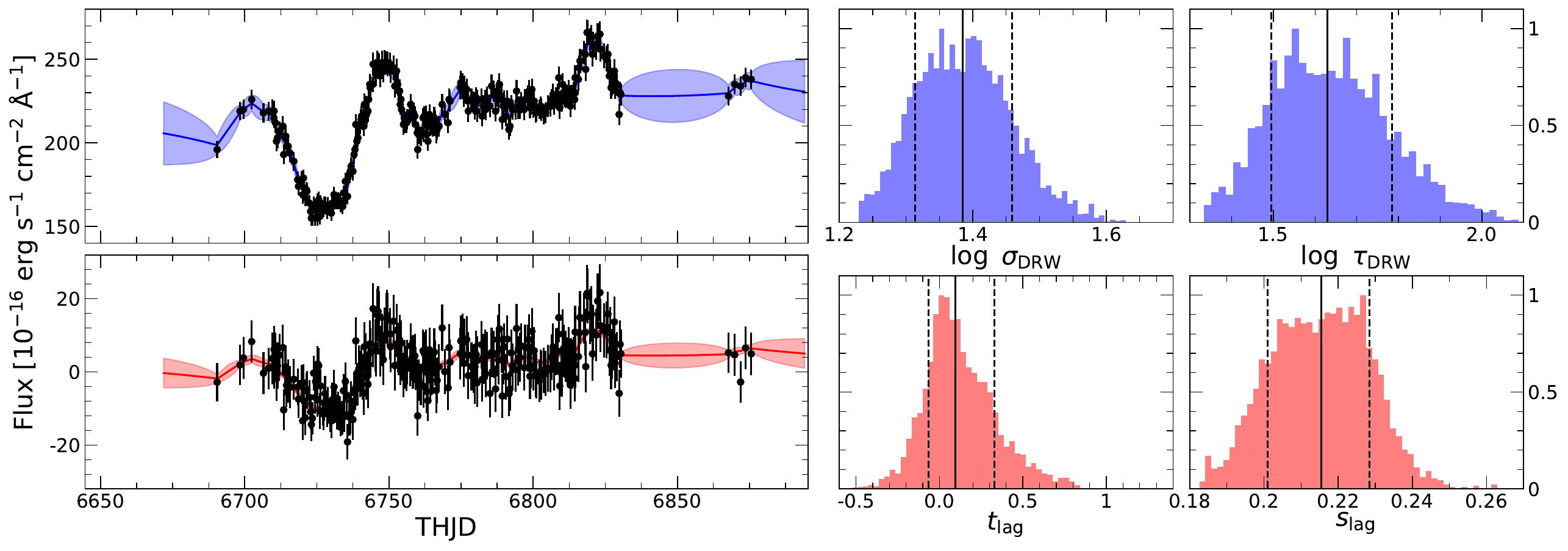}
\caption{\textbf{Left:} \javelin\ model fits to the unmodified $U$ lightcurve (\textbf{top}) and its corresponding residuals (\textbf{bottom}) for NGC~5548. \textbf{Right:} Posterior distribution for the \javelin\ parameter fits.  The recovered lag is consistent with 0~d, which is again peculiar if this is extra flux due to BLR contamination.}
\label{fig:ngc5548_resid_javelin}
\end{figure*}

Finally, we repeat our analysis using residual lightcurves without an inserted BLR signal for the AGN STORM~1 lightcurves of NGC~5548, which were initially analyzed in NK22.  Before doing this, we insert a BLR signal into the NGC~5548 data and, similar to our tests of Mrk~817, we find negligible change to the temperature fluctuation maps along with a worsening of the goodness of fit. We are also able to recover the inserted BLR signal from the residuals with \javelin.   The \javelin\ fits for the residuals of the unmodified $U$ lightcurve are shown in Figure~\ref{fig:ngc5548_resid_javelin}.  We find a lag $t_{\rm lag} = 0.11^{+0.36}_{-0.17}$~days and scale $s_{\rm lag} = 0.22^{+0.02}_{-0.02}$, which are very similar to the Mrk~817 results and not consistent with a lag expected for BLR contamination.

%%%%%%%%%%%%%%%%%%%%%%%%%%%%%%%%%%%%%%%%%%%%%%%%%%%%%%%%%%%%%%%%%%%%%%%%%

\section{Discussion}\label{sec:discussion}

We analyze the AGN STORM~2 lightcurves of Mrk~817 to produce maps of the temperature fluctuations on the disk.  In Section~\ref{sec:results}, we find that the temperature fluctuations are dominated by coherent radial structures that move slowly ($v \ll c$) inwards and outwards in the disk.  These are in strong conflict with the idea that a central lamppost is the only source of the variability in the disk through reverberation, where fluctuations would only appear as structures moving outward at roughly the speed of light. This is consistent with the results for the other AGNs analyzed in NK22 and \citet{stone23}.  We find that the timescales associated with the temperature fluctuations do not strongly depend on radius -- e.g., the inner radii probed by our model are not significantly more variable than the outer radii -- and we estimate this timescale to be of order 100~days. 

In a \citet{shakura73} thin disk model, the thermal and viscous timescales depend as $R^{3/2}$, and thus the weak radial dependence on the thermal fluctuations is difficult to explain without invoking a more complicated disk.  For example, a disk with a scale height that increases with radius would have variability timescales less dependent on radius (see e.g., \citealt{yao23}).  Alternatively, the temperature profile of the disk could be steeper than $T(R) \propto R^{-3/4}$, and so the range of radii being probed by our bands could be much narrower than what we see in Figure~\ref{fig:kernels}.  This would mean that our maps are probing a smaller range of radii that would all have similar timescales for variability.  Finally, if the thermal fluctuations are opacity-driven convection currents like those seen in accretion disk simulations \citep{jiang19,jiang20}, then the timescales of variability no longer depend on radius but instead only on mass and accretion rate.  While our model initially assumes a thin disk, as we discuss in Section~\ref{sec:methods}, none of these modifications would significantly impact the qualitative structures we observe in our maps. They would only shift, shrink, or stretch the radii where the $\delta T$ fluctuations exist.

In Section~\ref{sec:smooth}, we investigate how smoothing the lightcurves and then modeling them changes the resulting temperature maps.  We also examine how subtracting out these smoothed lightcurves changes the maps.  We find that smoothing the lightcurves in time does not lead to a change in the structure of the temperature fluctuations. A key insight from this exercise is that the temperature fluctuations produce effects on the lightcurves that exist over relatively long timescales (> 56~days).  These effects are often removed by detrending the lightcurves in order to more cleanly detect the reverberation signal (e.g., \citealt{mchardy14,mchardy18,pahari20,cackett23,miller23}).  In other words, the signal that is excluded by detrending is being produced by the slow-moving temperature fluctuations that we see in our maps.  This can explain some of the issues with measuring disk and UV broad emission line reverberation signals that are discussed in \citet{cackett23} and \citet{homayouni23b}.  However, it is worth noting that both papers explore alternative explanations invoking variable obscuration which we do not consider.

\begin{figure*}
\centering
\includegraphics[width=0.85\linewidth]{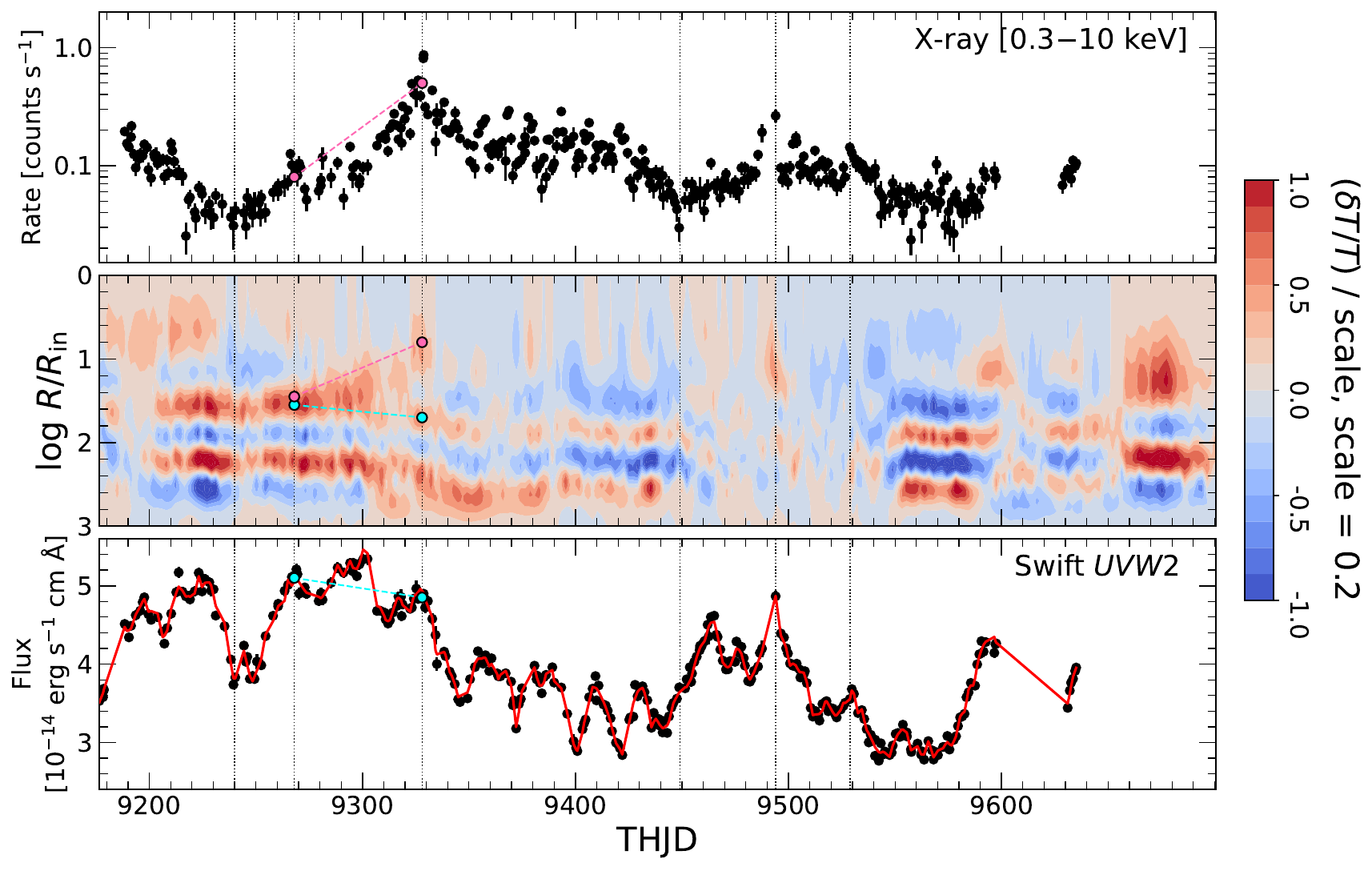}
\caption{\textbf{Top:} Swift XRT 0.3--10~keV lightcurve in observed counts.  \textbf{Center:} Temperature map ($\xi = 10$).  Note that the y-axis has been inverted (inner radii are at the top, outer at the bottom) compared to earlier figures.  \textbf{Bottom:} Swift~$UVW2$ lightcurve (black) and model fits (red).  Dashed black lines correspond to noticeable features in the X-ray and UV lightcurves.  The pink and cyan lines in the temperature map seem to correspond to similar structures in the X-ray and UV lightcurves, respectively. }
\label{fig:mrk817_xray}
\end{figure*}

In Section~\ref{sec:rm}, we test how contamination of the lightcurves by the BLR continuum emission can affect our temperature maps by inserting artificial signals meant to mimic such contamination.  We find that our model is unable to reproduce the artificial signals.  This is because the model is unable to recreate a signal restricted to a limited wavelength range due to the large overlaps in the disk radii contributing to any band (see Fig.~\ref{fig:kernels}) -- a disk with a smooth emission profile cannot produce a signal localized in wavelength. This problem does not change with a more realistic model for the BLR contamination (see Fig.~\ref{fig:mrk817_netzer}) because the signal in the $U$ and $u$ bands is still significantly larger than in adjacent bands, though this could change by using a different model for BLR emission with perhaps a ``smooth'' Balmer break that is indicative of strong turbulence in the BLR gas \citep{korista19,netzer22}.

If we model the residuals from the lightcurves with the artificial BLR signal with \javelin\ -- i.e., treat the residuals as an RM observation -- we recover the lag and amplitude of the inserted signal.  However, if we model the residuals of the unmodified $U$ lightcurve, we find they are consistent with a lag of 0~days for both Mrk~817 and NGC~5548, which is not consistent with BLR continuum contamination where we should obtain a lag time of order the BLR light travel time.  A possible explanation for these residuals is that the emission is reprocessed emission from the UV/X-ray obscurer discussed in \citet{kara21}, \citet{homayouni23a}, and \citet{partington23}. This obscurer is thought to be situated in the inner BLR or further inwards.  If the obscurer contributes significant Balmer continuum flux, it would result in a shorter lag than the BLR, perhaps even the 0-day lag we see in our \javelin\ results.  However, if reprocessed emission from the obscurer is significant, then we might also expect to see other emission features, like anomalously broad Balmer emission lines.  However, such lines are not observed in the spectra of Mrk~817.   Interestingly, the lag spectrum of Mrk~817 does not show an extra lag ``bump'' in the $U$ and $u$ bands after detrending \citep{cackett23}, whereas NGC~5548 had such a bump \citep{fausnaugh16}.  Yet, the results of our analysis yield the same scale of fluctuations and same lag time of 0~days.  In any case, the residual flux does not appear reproducible with our disk model, and so it is possible that the residual flux is coming from a non-disk component like the BLR.  This non-disk component is limited to contribute only $\sim$20\% of the variable flux in the $U$ and $u$ band lightcurves.  

Our analysis of the residuals from our model to search for BLR contamination is an example of our model being used in a \textit{predictive} capacity rather than \textit{descriptive}.  In addition to producing maps themselves, this model can be used to try to analyze other aspects of AGN variability that are not directly probed by the model.  In Figure~\ref{fig:mrk817_xray}, we show the 0.3--10~keV Swift~XRT lightcurve, along with a UV lightcurve and the $\xi = 10$ temperature map.  We highlight several features of the X-ray lightcurve which arguably match up with similar features in the UV lightcurve.  However, one key distinction in the X-ray flare/maximum at THJD~9328.  This flare was characterized in \citet{partington23} as a relatively small change in the intrinsic X-ray flux combined with a large change in the ionization state of the obscuring gas. While this strong X-ray flare corresponds to a small flare in the UV, it does not correspond to the UV maximum, which occurs about $\sim$30~days earlier.  In the lead up to the X-ray flare, there is a positive temperature fluctuation that appears to move inwards towards $\log R/R_{\rm in} = 0$.  At the same time, there is also a perturbation that moves slowly outwards, following the apparent motions discussed in Section~\ref{sec:results}.  These two fluctuations seem to track the X-ray and UV variability, respectively, where the inward fluctuation tracks the rising X-ray flux, and the outward fluctuation tracks the declining UV.  These associations between disk structures and lightcurve behaviors are tenuous, but seem interesting, especially because the model knows \textit{absolutely nothing} about the X-ray lightcurve.  Any association between the two can then imply a physical connection between the two, and perhaps a ``solution'' to the long-standing problem of uniting the UV disk and X-ray corona.

We are still unsure of what physical process is generating the slowly-moving $\delta T$ fluctuations in the disk, but based on the timescales involved and their presence in other AGNs \citep{neustadt22,stone23}, it is likely that these fluctuations emerge from variability mechanisms intrinsic to the accretion disk itself.  Advances in accretion disk simulations are thus needed to identify the physical mechanisms that generate the structures in our $\delta T$ maps.  Clearly, high-cadence, multi-band lightcurves like those obtained for the AGN~STORM campaigns are vital for characterizing this aspect of disk variability.  While the wavelength range will be more limited than the AGN~STORM campaigns, the upcoming Vera Rubin Observatory/LSST \citep{ivezic19} will also provide a unique opportunity to perform a large-scale analysis on millions of AGN lightcurves.  The long baseline and near-daily-cadence will be especially important in this regard, as it will allow a better characterization of the timescales of the temperature fluctuations, and of long-timescale AGN variability in general. 

\facilities{HST (COS), Swift, LCO, Liverpool:2 m, Wise Observatory, Zowada, CAO:2.2 m, YAO:2.4 m}

%%%%%%%%%%%%%%%%%%%%%%%%%%%%%%%%%%%%%%%%%%%%%%%%%%%%%%%%%%%%%%%%%%%%%%%%%

\section*{Acknowledgements}

Our project began with the successful Cycle 28 HST proposal 16196 \citep{peterson20}.  Support for Hubble Space Telescope program GO-16196 was provided by NASA through a grant from the Space Telescope Science Institute, which is operated by the Association of Universities for Research in Astronomy, Inc., under NASA contract NAS5-26555. 

J.M.M.N. thanks Z.~Yu and N.~Downing for assistance with \javelin.  J.M.M.N. and C.S.K. are supported by NSF grants AST-1814440 and AST-1908570.  C.S.K. is supported by NSF grant AST-2307385.  J.G. gratefully acknowledges support from NASA through grant 80NSSC22K1492.  Research at UC Irvine was supported by NSF grant AST-1907290.  E.M.C. gratefully acknowledges support from NASA through grant 80NSSC22K0089 and support from the NSF through grant No. AST-1909199. H.L. acknowledges a Daphne Jackson Fellowship sponsored by the Science and Technology Facilities Council (STFC), UK.  M.C.B. gratefully acknowledges support from the NSF through grant AST-2009230.  A.V.F. is grateful for financial assistance from the Christopher R. Redlich Fund and numerous individual donors. Y.H. was supported as an Eberly Research Fellow by the Eberly College of Science at the Pennsylvania State University.  Y.H. acknowledges support from the Hubble Space Telescope program GO-16196, provided by NASA through a grant from the Space Telescope Science Institute, which is operated by the Association of Universities for Research in Astronomy, Inc., under NASA contract NAS5-26555.  D.I., A.B.K, and L.\v C.P. acknowledge funding provided by the University of Belgrade - Faculty of Mathematics (the contract 451-03-68/2022-14/200104), Astronomical Observatory Belgrade (the contract 451-03-68/2022-14/ 200002), through the grants by the Ministry of Education, Science, and Technological Development of the Republic of Serbia.  D.I. acknowledges the support of the Alexander von Humboldt Foundation.  A.B.K. and L.{\v C}.P thank the support by Chinese Academy of Sciences President's International Fellowship Initiative (PIFI) for visiting scientist.  Y.R.L. acknowledges financial support from NSFC through grant Nos. 11922304 and 12273041 and from the Youth Innovation Promotion Association CAS.  M.R.S. is supported by the STScI Postdoctoral Fellowship.  M.V. gratefully acknowledges support from the Independent Research Fund Denmark via grant number DFF 8021-00130. This work made use of data supplied by the UK Swift Science Data Centre at the University of Leicester.

%%%%%%%%%%%%%%%%%%%%%%%%%%%%%%%%%%%%%%%%%%%%%%%%%%%%%%%%%%%%%%%%%%%%%%%%%
\bibliography{bibliography}{}
\bibliographystyle{aasjournal}
%%%%%%%%%%%%%%%%%%%%%%%%%%%%%%%%%%%%%%%%%%%%%%%%%%%%%%%%%%%%%%%%%%%%%%%%%

%%%%%%%%%%%%%%%%%%%%%%%%%%%%%%%%%%%%%%%%%%%%%%%%%%%%%%%%%%%%%%%%%%%%%%%%%

\appendix 
\section{Notes on picking the ideal penalty factor}\label{sec:append}

In NK22, the most important equations for our model are the following.  We start with 

\begin{equation}
\delta F = W \delta T
\end{equation}
where $\delta F$ are the lightcurve fluxes, $\delta T$ are the temperature fluctuations, and $W$ is the system of equations that relates the two quantities.  Using linear regularization, we invert the system of equations to find

\begin{equation}
\delta T  =\Big[W_\sigma^T W_\sigma + \xi(I_T + D_{kT} + D_{lT}) \Big]^{-1} W^T_\sigma \delta F_\sigma  ~ 
\end{equation}
where the $\sigma$ subscripts denotes normalizing for (dividing by) the errors $\sigma$, and the terms being multiplied by the penalty factor $\xi$ are the regularization terms that try to minimize the scale of the fractional temperature fluctuations $\delta T/T$ and large variations in $\delta T/T$ between adjacent temporal and radial grid elements.  For the definitions of these matrices, see Section~2 of NK22.  

The degrees of freedom $\nu$ for the linear regularization are defined (see, e.g., \citealt{ivezic20}) as
\begin{equation}
\nu  = {\rm Tr} \Big( W_\sigma \Big[W_\sigma^T W_\sigma + \xi(I_T + D_{kT} + D_{lT}) \Big]^{-1} W^T_\sigma  \Big) ~.
\end{equation}
The degrees of freedom thus depends on the errors and the penalty factor, roughly with the structure $\nu \propto (1+\sigma^2\xi)^{-1}$ so that $\nu$ decreases as the smoothing increases.  For $\xi = 1, 10, 100, 1000$, we get $\nu = 1378, 1064, 791, 530$, respectively.  From this, we get $\chi^2_\nu = 8.60, 12.26, 18.14, 31.37$, respectively, which increases more rapidly with $\xi$ than the $\chi^2/N_d$ (see Fig.~\ref{fig:mrk817_model}), but $\chi^2_\nu$ is generally not used as a metric for the quality of the model in linear regularization problems.  

Two metrics that are used are the Bayesian and Akaike information criteria (BIC and AIC, respectively).  For linear regularization, the BIC and AIC are formulated as
\begin{equation}
\begin{split}
{\rm AIC} & ~= \chi^2 + 2\nu ~, \\ 
{\rm BIC} & ~= \chi^2 + \nu\ln{N_d}  ~, \\ 
\end{split}
\end{equation}
where the main difference is that the BIC penalizes having more parameters more than the AIC does since $\ln N_d > 2$.  To choose the best model using the AIC or BIC, one chooses the model that minimizes the selected criterion.  

\begin{figure*}[htb!]
\centering
\includegraphics[width=0.47\linewidth]{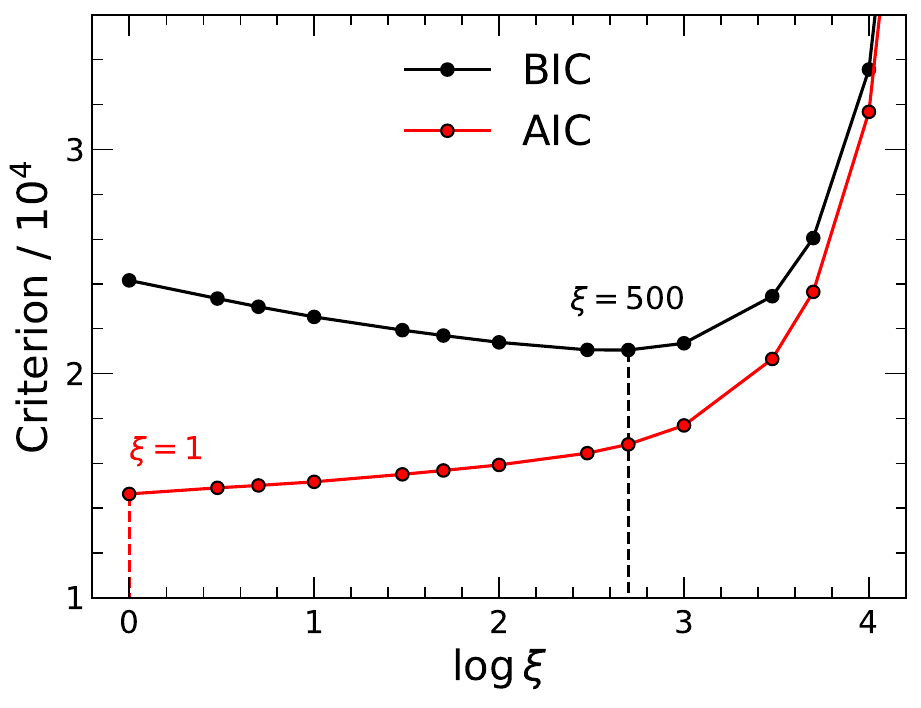}
\caption{AIC and BIC as functions of the smoothing parameter $\xi$.  The ``ideal'' smoothing parameters that minimize the AIC and BIC are marked.  The two criteria prefer very different parameters which roughly span the $\delta T$ maps shown in Fig.~\ref{fig:mrk817_model}.}
\label{fig:mrk817_bic}
\end{figure*}

We show the two criteria as a function of $\xi$ for our model in Figure~\ref{fig:mrk817_bic}.  The minimum values of the AIC and BIC correspond to $\xi \simeq 1$ and $\xi \simeq 500$, respectively.  It is possible that the AIC prefers a value lower than $\xi = 1$, but as we discuss in Section~\ref{sec:methods}, a small $\xi$ leads to unphysically high fractional fluctuations in $\delta T/T$. While the two criteria favor very different smoothing parameters, the range spanned by these two values roughly corresponds to the maps shown in Figure~\ref{fig:mrk817_model}.  We again note that the qualitative structures of the patterns are roughly the same between $\xi=1$ and $\xi=100$, and while for $\xi=1000$ the radial structures in the maps are strongly suppressed, they are arguably still present.  We have no reason to favor one criterion over the other, but we do also have a physical ``prior'' -- our equations assume that $\delta T$ behaves linearly, and so we need to keep $\delta T/T$ in the linear regime ($\lesssim 0.2$).  This seems to be as good as any metric for focusing on the $\xi=10$ solution in Figure~\ref{fig:mrk817_model_zoom}.

% One approach is to evaluate the ``L-curve,'' so called because of it typically resembling the L-shape.  The curve plots $\log \chi^2$ against $\log H$ as a function of $\xi$.  The ideal $\xi$ value is located at the inflection point of the L-curve where, as $\xi$ increases, the curve switches from the regime where $H$ is rapidly decreasing (and $\chi^2$ is slowly increasing) to a regime where $\chi^2$ is rapidly decreasing (and $H$ is slowly decreasing).

%DOF for xi = 1,10,100,1000 are 1383.0, 1068.5, 795.2 and 536.6, 
%%%%%%%%%%%%%%%%%%%%%%%%%%%%%%%%%%%%%%%%%%%%%%%%%%%%%%%%%%%%%%%%%%%%%%%%%

\end{document}